\documentclass[11pt,letterpaper]{article}

\usepackage[T1]{fontenc}
\usepackage[utf8]{inputenc}
\usepackage{lmodern}
\usepackage[margin=0.95in]{geometry}
\usepackage{microtype}
\usepackage{amsmath,amssymb}
\usepackage{graphicx}
\usepackage[table]{xcolor}
\usepackage{float}
\usepackage{multicol}
\usepackage{tabularx}
\usepackage{caption}
\usepackage{tikz}
\usepackage{pgfplots}
\usepackage{url}
\usepackage[hidelinks]{hyperref}
\pgfplotsset{compat=1.17}
\usetikzlibrary{shadows,positioning,backgrounds,arrows.meta,shapes.geometric,shapes.multipart,calc,fit}

\hypersetup{
  pdftitle={CyberHull: Deterministic Malicious URL Isolation via Convex Hull},
  pdfauthor={Avijit Gayen, Aniket Ghosal, Durba Chakraborty, Sayan Mondal, Md Abdul Aziz Al Aman, Angshuman Jana}
}
\renewcommand{\arraystretch}{1.1}

\title{${\mathbb{C}}$yber${\mathbb{H}}$ull: Deterministic Malicious URL Isolation via Convex Hull}
\author{\begin{tabular}{c}
Avijit Gayen, Aniket Ghosal, Durba Chakraborty,\\
Sayan Mondal, Md Abdul Aziz Al Aman, Angshuman Jana\\[0.6em]
{\small Department of CSE, Techno India University, Kolkata, India}\\
{\small Department of CSE (AI), Techno India University, Kolkata, India}\\
{\small Department of CSE, Indian Institute of Information Technology, Assam, India}\\[0.45em]
{\footnotesize Emails: \href{mailto:avijit.gayen@iiitg.ac.in}{\texttt{avijit.gayen@iiitg.ac.in}},}\\
{\footnotesize \href{mailto:sayan.mondal@iiitg.ac.in}{\texttt{sayan.mondal@iiitg.ac.in}}, \href{mailto:angshuman@iiitg.ac.in}{\texttt{angshuman@iiitg.ac.in}}}
\end{tabular}}
\date{}
\begin{document}
\maketitle
\begin{abstract}
The emergence of increasingly sophisticated and evasion-resistant phishing attack strategies represents a key risk to today's cybersecurity landscape involving cyber-physical systems and internet-based networks. Traditional probabilistic methods and some forms of machine learning have difficulty dealing with novel attacks and may lack interpretability or portability. To address this challenge, we present $\mathbb{C}$yber$\mathbb{H}$ull, a resource-efficient and deterministic approach that identifies malicious URLs through the formation of mathematically defined convex hull perimeters across multiple orthogonal feature domains (structural content entropy, hyperlink graph topography, semantic token distribution). $\mathbb{C}$yber$\mathbb{H}$ull trains relatively small convex hulls known as “vaults” using QuickHull algorithm (complexity of $\mathcal{O}(n \log n)$ during training stage), while each point-in-polyhedron test of a request entails low computational overhead on standard CPUs. Our experiment demonstrates that $\mathbb{C}$yber$\mathbb{H}$ull achieves 92\% detection accuracy and significantly surpasses a basic linear approach based on PCA algorithm (32\%). Moreover, this method is computionally more efficient compared to contemporary deep learning solutions.
\end{abstract}

\section{Introduction} \label{Sec: Introduction}
The phishing site can be described as a website that mimics the websites of well-known organizations. It is intended to trick users into surrendering their login information and other sensitive data. The use of such websites as a launching pad for cyber attacks means that phishing pages have become more and more sophisticated. The use of advanced phishing frameworks is an emerging and serious threat to the worldwide network systems.

% Domain Motivation
% Today’s cyber attackers no longer rely on simple fake domain names.  Instead, they use a pipeline process automatically generating URLs that use zero-day evasion techniques such as highly obfuscated HTML formats~\cite{hannousse2021web}, random routing~\cite{tamal2024comprehensive}, and disordered entropy levels in their content~\cite{sharma2023information}. These new phishing techniques have been purposely designed to mimic legitimate website activities and constantly change their online signatures making traditional blacklists and signature heuristic methods useless~\cite{patel2024topological}. Today, these evasion techniques cost the global economy billions of dollars per year in financial losses by successfully circumventing traditional perimeter defenses and severely compromising the integrity of critical cyber-physical systems.
Today’s cyber attackers have moved beyond simple fake domain names. Instead, they use automated phishing generation pipelines that can produce adaptive malicious URLs with advanced evasion techniques such as highly obfuscated HTML structures~\cite{hannousse2021web}, randomized routing mechanisms~\cite{tamal2024comprehensive}, and irregular entropy distributions in web content~\cite{sharma2023information}. Industry reports from recent years illustrate the scale and impact of these threats. The Anti-Phishing Working Group (APWG) has reported more than one million phishing campaigns each quarter~\cite{apwg_report_year} . The FBI Internet Crime Complaint Center (IC3) reports billions of dollars in annual losses due to phishing and business-email-compromise incidents~\cite{ic3_report_year}. By design, these adversaries mimic legitimate site behavior while constantly changing observable signatures, severely degrading the effectiveness of traditional blacklist and signature-based detection systems~\cite{patel2024topological}. This has resulted in the emergence of novel, previously unseen phishing campaigns that threaten the operational integrity of large-scale Internet infrastructures and critical cyber-physical systems.

% Motivation from the Literature
Therefore, it is vital to adopt a radical new approach from rule-based methods to structural anomaly detection processes in response to such a daunting situation. While the academic literature provides a variety of analytical countermeasures to address these challenges, all existing paradigms suffer from fundamental weaknesses. 

% ---\textcolor{blue}{This paragraph can be treated as general background and be kept small in size. Next paragraph should be motivation and it would be in two part: a)Domain motivation which would be supported by statistics and b) Motivation from the literature.--- I suggest divide the first two paragraphs into three small and coherent paragraphs.}

% Limitations of Existing Paradigms
To deal with such challenges, cybersecurity professionals have introduced different analytical methods~\cite{mahdavifar2019application}, but all of them involve essential weaknesses. Traditional statistical and probabilistic algorithms are usually ineffective since they operate under fixed constraints regarding past data distribution patterns~\cite{rosner1999use}; these systems are prone to quickly deteriorate in case if the attack causes changes to the nature of the signature left by the adversary. Graph-theoretic and heuristic methods, in turn, try to understand the network connections~\cite{gayen2024network} and structure of the content~\cite{tan2020graph}, but they are costly and can easily be interfered with by "noise," such as injecting innocent keywords and dead links by the attacker. Thus, DL systems are widely used because of their high performance in classifying different threats~\cite{vijayalakshmi2020web}. Nevertheless, the modern neural nets also pose certain issues since they are unable to provide deterministic explanations (see the table~\ref{tab: Capability Matrix}) . Namely, their training process is quite demanding in terms of computer power, and they are based on obscure parameter weights that do not allow one to determine why the algorithm classified certain domains as malicious in mathematical terms~\cite{chen2022constraints}.

\begin{table}[htbp]
\centering
\small
\caption{Technological Capability Matrix}
\label{tab: Capability Matrix}
% The resizebox forces the table to fit exactly within the column width
\resizebox{\columnwidth}{!}{%
\begin{tabular}{|l|c|c|c|}
\hline
\rowcolor{black!10} % Adds a light gray background to the header row
\textbf{Requirement} & \textbf{Statistical} & \textbf{DL} & \textbf{$\mathbb{C}$yber$\mathbb{H}$ull} \\
\hline
Previously Unseen Detection & \textcolor{red}{$\times$} & \textcolor{green!60!black}{$\checkmark$} & \textcolor{green!60!black}{$\checkmark$} \\
\hline
Explainability & \textcolor{green!60!black}{$\checkmark$} & \textcolor{red}{$\times$} & \textcolor{green!60!black}{$\checkmark$} \\
\hline
Edge Efficiency & \textcolor{green!60!black}{$\checkmark$} & \textcolor{red}{$\times$} & \textcolor{green!60!black}{$\checkmark$} \\
\hline
Autonomous Scaling & \textcolor{red}{$\times$} & \textcolor{green!60!black}{$\checkmark$} & \textcolor{green!60!black}{$\checkmark$} \\
\hline
\end{tabular}%
}
\end{table}

This paper contributes mainly by introducing $\mathbb{C}$yber$\mathbb{H}$ull, a deterministic spatial threat-isolation framework built on computational geometry, to address limitations in probabilistic classification and the lack of interpretability of deep learning-based threat detection systems. $\mathbb{C}$yber$\mathbb{H}$ull does not model dynamic adversarial behaviors in a constant manner, but instead uses an explicit verification paradigm, requiring each web request to meet pre-defined mathematical boundary conditions to be legitimate. The proposed framework uses convex geometric representations to give mathematically provable safe regions that represent predictable benign web traffic, enabling deterministic isolation of anomalous requests that stray outside the learned spatial boundaries.

% Methodology, Performance, and Implementation Challenges
% ---\textcolor{blue}{Next two paragraphs outline the major cross validation and performance of proposed model---I suggest that it would be part of methodology paragraph and be written very precisely. We need to report these in the following sections. Additionally, we need to report the implementation challenges of this model at the end of the methodology paragraph.} 
To this end, $\mathbb{C}$yber$\mathbb{H}$ull maps high-variance web attributes such as aggregate content entropy, semantic keyword distributions and hyperlink topology ratios into a multi-dimensional coordinate space. The framework then constructs several independent geometric “Vaults” for cross-space verification using QuickHull~\cite{barber1996quickhull}-based convex boundary generation. Thus, even if a sophisticated phishing payload succeeds in mimicking one structural property of legitimate traffic, it is mathematically bound to violate the orthogonal constraints of another Vault during spatial containment verification~\cite{patel2024topological}. This multi-tier geometric verification strategy also enhances interpretability by substituting opaque probabilistic decision boundaries with mathematically explainable spatial constraints. Experiments on benchmark phishing datasets demonstrate that the proposed multi-tier geometric framework achieves 92\% detection accuracy with significantly lower computational overhead compared to deep learning alternatives. However, applying the framework in practice poses several challenges, including the computational cost of mapping high-dimensional spaces and the need to periodically re-calibrate convex boundaries to account for the evolving topology of legitimate web infrastructures over time.

% % Dataset & Empirical Preview
% To confirm the effectiveness of the geometric model, $\mathbb{C}$yber$\mathbb{H}$ull was experimentally tested on an extensive benchmark database including both legitimate web traffic and sophisticated zero-day phishing URLs~\cite{tamal2024comprehensive}. In total, 91 different multidimensional features were collected from this sample, ranging from structural entropy to network hyperlink percentages to semantic keyword frequencies~\cite{hannousse2021web, sharma2023information}. With the help of these concrete samples being mapped to the geometric Vaults, the system efficiently encompassed the mathematical heart of legitimate traffic with great accuracy. The test results have confirmed that the proposed zero-trust geometric framework was able to accurately detect phishing attacks at a rate of 92\%. More importantly, the algorithm managed to capture adversaries that had evaded traditional linear classifiers, showing that pure spatial mathematics is capable of delivering precise real-time detection without the heavy computational power required by deep learning systems~\cite{chen2022constraints}.

% Key Contributions \& The Autonomous controlled Pipeline explanation
Expanding upon the above empirical results, this study will tackle the weaknesses inherent to probabilistic threat detection by way of the following fundamental innovations:
\begin{itemize}
    \item  We propose $\mathbb{C}$yber$\mathbb{H}$ull, a deterministic geometric threat-isolation framework, leveraging Convex Hulls to derive mathematically verifiable spatial boundaries for legitimate web traffic.
    \item A multi-tier cross-reference geometric verification strategy for detecting adversarial phishing payloads across multiple orthogonal feature spaces, and achieving 92\% detection accuracy with a significantly lower computational overhead than conventional deep learning-based approaches.
\end{itemize}

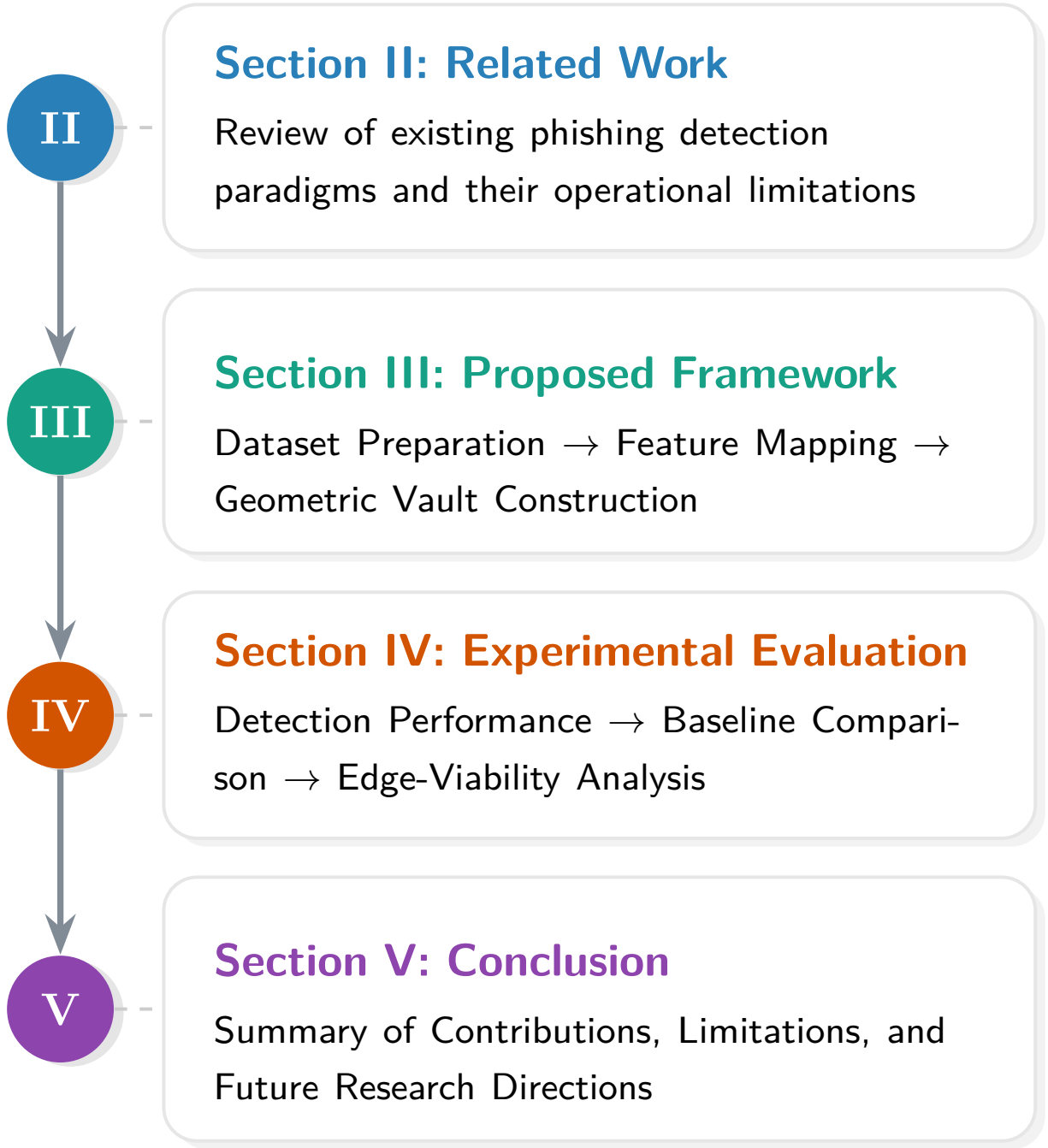
\begin{figure}[!t]
    \centering
    % Scales the figure to perfectly fit a single column width
    \resizebox{0.95\columnwidth}{!}{%
    \begin{tikzpicture}[
        font=\sffamily,
        node distance=0.3cm,
        % Style for the text boxes
        box/.style={
            rectangle, draw=gray!20, fill=white, thick, 
            rounded corners=8pt, text width=6.5cm, align=left, 
            inner xsep=12pt, inner ysep=10pt, drop shadow={opacity=0.1, shadow yshift=-2pt}
        },
        % Style for the Colored Section Roman Numerals
        sec_num/.style={
            circle, fill=#1, text=white, font=\bfseries, 
            inner sep=0pt, minimum size=0.9cm,
            drop shadow={opacity=0.2, shadow yshift=-1pt}
        }
    ]
    
    % Define Premium Vibrant Colors
    \definecolor{col2}{HTML}{2980B9} % Professional Blue
    \definecolor{col3}{HTML}{16A085} % Modern Teal
    \definecolor{col4}{HTML}{D35400} % Energetic Orange
    \definecolor{col5}{HTML}{8E44AD} % Deep Purple
    \definecolor{slate}{HTML}{2C3E50} % Dark Slate for lines

    % --- SECTION II ---
    \node[box] (sec2) {
        \textcolor{col2}{\textbf{Section II: Related Work}}\\[0.5ex]
        \footnotesize Review of existing phishing detection paradigms and their operational limitations
    };
    \node[sec_num=col2, left=0.4cm of sec2.west, anchor=east] (n2) {II};
    \draw[thick, gray!40, dashed] (n2) -- (sec2.west);

    % --- SECTION III ---
    \node[box, below=of sec2] (sec3) {
        % \textcolor{col3}{\textbf{Section III: Proposed Model}}\\[0.5ex]
        % \footnotesize 
        % $\bullet$ Benchmark Dataset Description \& Setup\\[0.5ex]
        % $\bullet$ Probabilistic Vault Selection \& Spatial Mapping\\[0.5ex]
        % $\bullet$ The $\mathbb{C}$yber$\mathbb{H}$ull Methodology

        \textcolor{col3}{\textbf{Section III: Proposed Framework}}\\[0.5ex]
        \footnotesize 
        Dataset Preparation $\rightarrow$ Feature Mapping $\rightarrow$ Geometric Vault Construction
    };
    \node[sec_num=col3, left=0.4cm of sec3.west, anchor=east] (n3) {III};
    \draw[thick, gray!40, dashed] (n3) -- (sec3.west);

    % --- SECTION IV ---
    \node[box, below=of sec3] (sec4) {
        % \textcolor{col4}{\textbf{Section IV: Results \& Discussion}}\\[0.5ex]
        % \footnotesize 
        % $\bullet$ Global Detection Efficacy vs. Linear Baselines\\[0.5ex]
        % $\bullet$ Multi-Dimensional Threat Isolation Analysis\\[0.5ex]
        % $\bullet$ Discussion \& Edge-Viability (Line-Rate Economics)
        \textcolor{col4}{\textbf{Section IV: Experimental Evaluation}}\\[0.5ex]
        \footnotesize 
        Detection Performance $\rightarrow$ Baseline Comparison $\rightarrow$ Edge-Viability Analysis
    };
    \node[sec_num=col4, left=0.4cm of sec4.west, anchor=east] (n4) {IV};
    \draw[thick, gray!40, dashed] (n4) -- (sec4.west);

    % --- SECTION V ---
    \node[box, below=of sec4] (sec5) {
        % \textcolor{col5}{\textbf{Section V: Conclusion}}\\[0.5ex]
        % \footnotesize 
        % $\bullet$ Summary of Core Contributions\\[0.5ex]
        % $\bullet$ Operational Limitations \& Future Scope

        \textcolor{col5}{\textbf{Section V: Conclusion}}\\[0.5ex]
        \footnotesize 
        Summary of Contributions, Limitations, and Future Research Directions
    };
    \node[sec_num=col5, left=0.4cm of sec5.west, anchor=east] (n5) {V};
    \draw[thick, gray!40, dashed] (n5) -- (sec5.west);

    % --- Connecting Vertical Arrows ---
    \draw[thick, slate!60, ->, >=Stealth, line width=1.5pt] (n2) -- (n3);
    \draw[thick, slate!60, ->, >=Stealth, line width=1.5pt] (n3) -- (n4);
    \draw[thick, slate!60, ->, >=Stealth, line width=1.5pt] (n4) -- (n5);

    \end{tikzpicture}%
    }
    \vspace{0.1cm}
    \caption{Structural organization and thematic flow of the remainder of the paper.}
    \label{fig:paper_org}
\end{figure}

% Most importantly, due to the fact that this geometric system does not rely on hidden weight parameters but rather utilizes purely mathematical spatial geometry, no heuristic calibration is needed. This system allows for native integration of complete artificial intelligence automation, providing the necessary theoretical basis for future machine-driven autonomous threat detection systems. In order to illustrate the importance of this spatial revolution, the subsequent sections will trace the history of modern threat detection techniques and examine how certain operational deficiencies in current machine learning algorithms and heuristics can be addressed by $\mathbb{C}$yber$\mathbb{H}$ull. ---\textcolor{red}{Make a small figure to explain the paper organization in short.}
Most importantly, the proposed geometric framework relies on the use of mathematically verifiable spatial representations, rather than on opaque hidden weight optimization, thus the decision-making process is inherently interpretable and computationally lightweight. Moreover, the deterministic nature of the framework reduces the need for extensive heuristic tuning, making it more suitable for scalable and automated threat analysis environments. The following sections present the evolution of modern phishing detection techniques and discuss the potential of overcoming the operational limitations of traditional machine learning and heuristic-based systems with the proposed $\mathbb{C}$yber$\mathbb{H}$ull framework to demonstrate the significance of this geometric detection paradigm. The structure of the paper is illustrated in Fig.~\ref{fig:paper_org}.

% --- SMALL PAPER ORGANIZATION FIGURE (Upgraded, Colorful & Classy) ---

\vspace{0.2cm} 
% -------------------------------------------------------

\vspace{1em}
%% =====================
\section{Related Work} \label{sec: Related Work}
% Learning-Based Approaches in Threat Detection
The trend within malicious URL detection is a departure from conventional blacklists to more sophisticated learning and behavioral approaches. Contemporary research into malicious URLs has relied significantly on natural language processing (NLP) and TF-IDF extraction to determine the semantics of the web content~\cite{sharma2023information, zhang2011understanding}. On that premise, various scholars have made use of ensemble learning and SVM techniques to predict malicious URLs through structural entropy and lexical attributes~\cite{vijayalakshmi2020web}. Recent studies have also looked into hybrid and ensemble-driven cybersecurity frameworks that combine statistical filtering, graph analytics and machine learning classifiers to achieve better robustness against adversarial manipulation and feature obfuscation~\cite{catal2022applications}. The goal of these approaches is to combine the interpretability of structured analytical methods with the adaptive learning capabilities of modern classifiers~\cite{arrieta2020explainable}. However, while these hybrid systems achieve better detection performance, they still largely rely on probabilistic inference mechanisms and often inherit the computational overhead and explainability limitations associated with conventional machine learning models. In recent times, deep neural networks~\cite{catal2022applications, mahdavifar2019application} and graph-based techniques~\cite{gayen2024network, tan2020graph} have been extensively explored. These paradigms exhibit exceptional ability to detect hyperlink connections and extract latent non-linear features, thereby enhancing classification performance.

However, despite these systems demonstrating impressive abilities, the contemporary literature identifies their significant drawbacks. For one, deep learning models require considerable amounts of time for training alongside intensive computer memory space~\cite{chen2022constraints}; hence, these systems cannot function efficiently within edge devices that are limited by computing power and storage capacity. Moreover, since such systems utilize probabilistic parameters for prediction rather than spatial rules, they fail to provide sufficient transparency on why a particular anomaly triggered an alarm~\cite{arrieta2020explainable}. This precise problem of lacking mathematical proof has compelled scholars in other critical areas to explore geometric models.

% Computational Geometry in High-Stakes Domains
% In addition to the computation-intensive deep learning models~\cite{chen2022constraints}, the classical methods of statistics and linear dimensionality reduction were also found ineffective for tackling the challenge posed by the domain of web-based attacks~\cite{patel2024topological}. The conventional ML process involves the utilization of techniques such as PCA for data reduction and boundary creation. Unfortunately, the linear transformation employed in PCA requires the input to follow a normal distribution and be linearly separable. The n-dimensional space of the current web-based threat landscape features complex and unpredictable entropy, semantic anomalies, and hyperlinks~\cite{sharma2023information}. To substantiate the ineffectiveness of PCA, we conducted an experiment where we applied the technique against our benchmark dataset~\cite{tamal2024comprehensive} comprising the entire corpus used for our geometric approach that resulted in a detection rate of 92\%. As expected, PCA produced an abysmally low separation accuracy of only 32\%, indicating that linear separation is impossible for such non-linear data.

Deep learning frameworks~\cite{chen2022constraints} have computational constraints, and classical statistical and linear dimensionality-reduction approaches have also exhibited important limitations in modern web-based threat environments~\cite{patel2024topological}. Popular machine learning techniques such as Principal Component Analysis (PCA) assume very strong assumptions on linear separability and normally distributed feature spaces~\cite{rosner1999use}. However, modern phishing ecosystems exhibit strongly non-linear characteristics with irregular entropy distributions~\cite{sharma2023information}, semantic anomalies and dynamically evolving hyperlink relationships~\cite{gayen2024network, tan2020graph}. Thus, simple linear transformation methods often do not yield reliable decision boundaries for complex adversarial web traffic patterns.

% The Research Gap (Positioning $\mathbb{C}$yber$\mathbb{H}$ull)
This practical problem in using linear classifiers, exacerbated by the "black box" nature of deep neural networks~\cite{catal2022applications, mahdavifar2019application, arrieta2020explainable}, has resulted in an important operational gap in the field of threat identification in real time. For applications in which predictability and transparency are key, scientists studying defensive mechanisms in physical spaces like autonomous UAV swarm tracking, missile trajectory representation, and radar border mapping have effectively eliminated probabilistic guesswork by employing computational geometry~\cite{barber1996quickhull}. In such high-stakes areas of study, N-dimensional Convex Hulls form inflexible and unforgiving spatial borders surrounding known objects. Nevertheless, although these robust mathematical concepts have found widespread use in the analysis of physical battlefields, they have seen little application in cybersecurity to date. Consequently, one major limitation is that no deterministic and geometrically meaningful cybersecurity approach has been developed yet, which can effectively separate malicious web traffic without depending solely on probabilistic learning models. $\mathbb{C}$yber$\mathbb{H}$ull is designed precisely for this purpose. This paper solves the problem by applying topological principles to map non-linear traits of the cyber domain into individual geometric Vaults~\cite{patel2024topological}.

%% =====================
\vspace{1em}
\section{Proposed Model} \label{Proposed Dataset}
In this section, we begin by describing the seed dataset utilized in this work.

\subsection{Benchmark Dataset Description} 
% Benchmark Dataset Description
In order to test the $\mathbb{C}$yber$\mathbb{H}$ull algorithm empirically, we used a data set based on the Web Page Phishing Detection Dataset~\cite{tamal2024comprehensive}. An initial seed of 11,430 URL samples, each representing 91 features, was put through an elaborate preprocessing procedure. In the end, the highly curated set consisted of 769 legitimate and 46 phishing URL samples, containing only those whose hyperlinks and semantic data could be completely extracted~\cite{hannousse2021web}. The drastic reduction in the size of the dataset was mainly due to the strict preprocessing constraints needed for deterministic geometric modelling. A significant proportion of the original URLs contained incomplete hyperlink structures, broken semantic metadata, inaccessible embedded resources, or partially missing entropy-related attributes, rendering them unsuitable for reliable high-dimensional spatial mapping. In the final experimental study, we considered only the samples with completely extractable and structurally valid feature spaces, because the proposed convex-boundary framework relies on mathematically consistent feature representations in all dimensions. The final dataset exhibits a very severe class imbalance between legitimate and phishing samples. This distribution was intentionally preserved. Operationally, this 16:1 ratio is an accurate approximation of real web environments and enables a more rigorous evaluation of the framework under real-world anomaly-isolation conditions. More important mathematically, this imbalance does not require any synthetic correction (like SMOTE). In contrast to probabilistic classifiers that minimize frequency-based loss functions, thus being sensitive to class imbalance, the $\mathbb{C}$yber$\mathbb{H}$ull framework depends only on spatial containment. A Convex Hull is defined only by the coordinates (vertices) of the outer boundary of the valid data cluster. Thus, the safe zone design is strictly invariant to the frequency distribution of malicious samples; hence, oversampling methods are mathematically irrelevant in the process of forming geometric boundaries. The final data set was presented in the form of a JSON object, containing deep attributes such as links distribution, term frequencies (TF-IDF)~\cite{zhang2011understanding}, and explicit Shannon entropy~\cite{sharma2023information}.
A cursory analysis of the data indicated specific architectural differences, supporting our choice of a geometric technique. Indeed, legitimate URLs create intricate graph structures~\cite{gayen2024network}, featuring intensive navigation within the site and numerous outbound links. Moreover, the semantics used in their descriptions include terms related to information (such as "institute" and "academic") and a high amount of structural entropy, implying diverse HTML content.
However, malicious domains have a very sparse and isolated network structure for hyperlinks that is characteristic of a domain meant only for credential harvesting~\cite{tan2020graph}. The semantic words in their case consist of an extremely high number of urgency-based words (such as login, secure, bank), leading to entropy reduction. Through the exploitation of such great differences in hyper-link density, intention, and entropy, we achieve a highly discriminating multi-dimensional basis in which our algorithm is evaluated.

\subsection{Probabilistic Vault Selection \& Spatial Mapping}

\begin{figure}[!ht]
    \centering
    \resizebox{0.95\textwidth}{!}{%
    \begin{tikzpicture}
        \begin{axis}[
            width=\textwidth,
            height=15cm, % Ensures all 46 bars are clearly visible
            xbar,
            bar width=5pt, 
            xmin=0, xmax=1.05,
            ymin=0, ymax=47,
            y dir=reverse, 
            xlabel={\textbf{Absolute Correlation Score (Mathematical Separability)}},
            ylabel={\textbf{Geometric Vault Rank}},
            ytick={1,2,...,46}, % Explicitly numbers every single vault
            yticklabel style={font=\scriptsize}, % Adjusts font size to prevent overlap
            ymajorgrids=true,
            grid style={dashed, gray!30},
            axis x line=bottom,
            axis y line=left,
            enlarge y limits=0.01
        ]
        
        % --- ALL 46 VAULTS (UNIFIED COLOR) ---
        % Using a premium, deep academic blue to represent that ALL 46 are selected and actively used.
        \addplot[fill=blue!80!black, draw=black!80, thick] coordinates {
            (0.99,1) (0.98,2) (0.97,3) (0.96,4) (0.95,5) (0.93,6) (0.92,7) (0.91,8) (0.88,9) (0.85,10) (0.83,11) (0.80,12) (0.78,13) (0.76,14) (0.75,15) (0.73,16) (0.72,17) (0.69,18) (0.66,19) (0.55,20) (0.53,21) (0.50,22) (0.47,23) (0.44,24) (0.37,25) (0.36,26) (0.35,27) (0.33,28) (0.31,29) (0.29,30) (0.28,31) (0.25,32) (0.24,33) (0.22,34) (0.20,35) (0.19,36) (0.18,37) (0.17,38) (0.16,39) (0.15,40) (0.14,41) (0.13,42) (0.12,43) (0.11,44) (0.10,45) (0.09,46)
        };
        
        \end{axis}
    \end{tikzpicture}
    }
    \vspace{-0.2cm}
    \caption{Ranking the 46 selected geometric vaults by absolute feature-pair separability score. The definitions of all vault feature pairs (V1--V46) are provided in Appendix~\ref{app:vault-pairs}.}
    \label{fig:vault_ranking}
\end{figure}
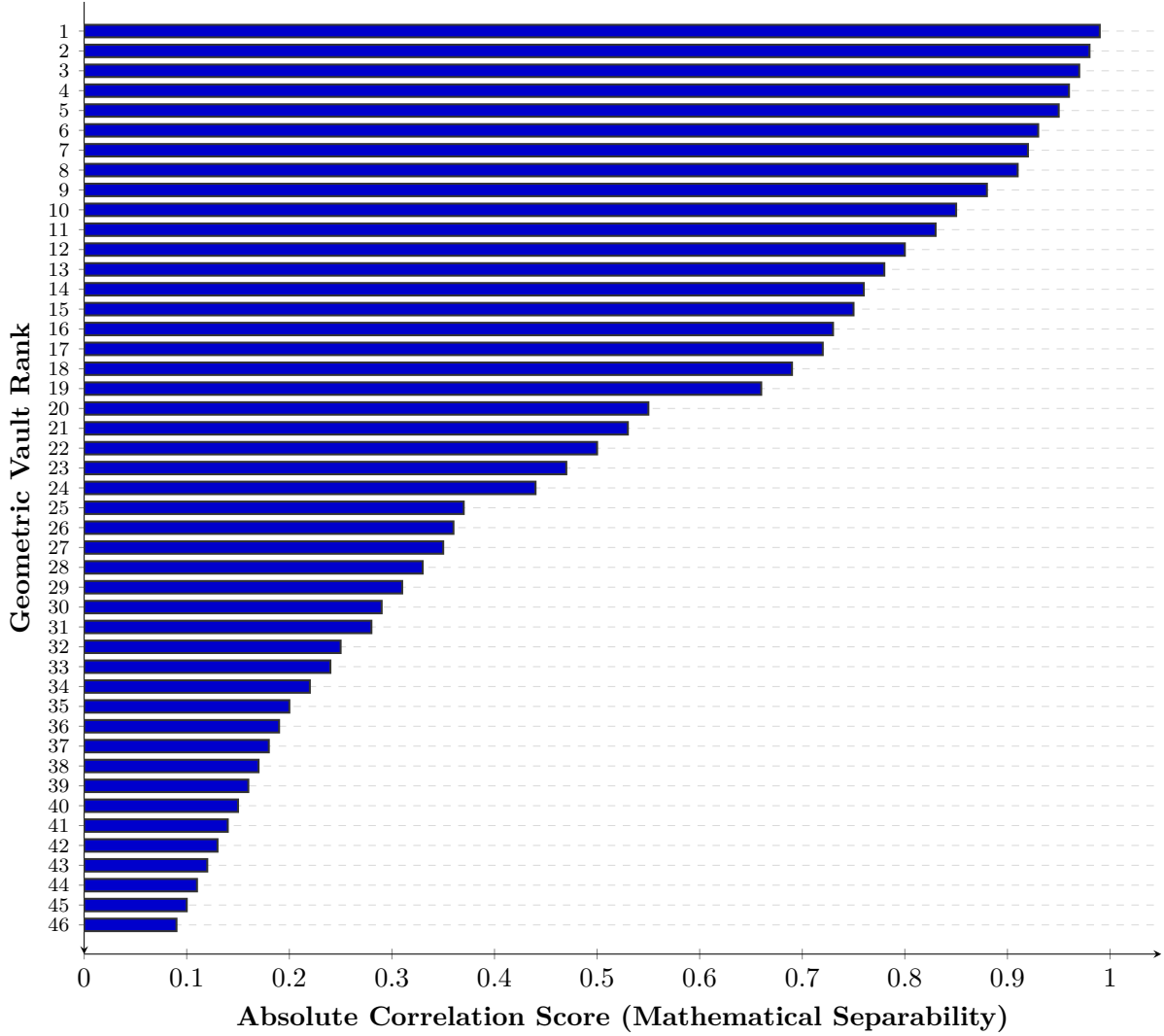

Unlike in heuristic approaches where the features chosen for use are rather arbitrarily picked, $\mathbb{C}$yber$\mathbb{H}$ull uses a mathematical method that determines the best coordinate axes to use for geometric bounding through a probabilistic approach~\cite{rosner1999use}. Let the set of features extracted be represented as $F=\{f_1, f_2, \dots, f_{91}\}$. For each such feature $f_i$, the probabilities $P(f_i\mid\text{Legitimate})$ and $P(f_i\mid\text{Malicious})$ are computed. Separability measure for every feature was calculated using the following normalized probabilistic distance metric:

\[
S(f_i)=\frac{\left|\mu_L(f_i)-\mu_M(f_i)\right|}{\sigma_L(f_i)+\sigma_M(f_i)}
\]

where $\mu_L(f_i)$ and $\mu_M(f_i)$ are the means for feature $f_i$ in the legitimate and malicious sample sets, and $\sigma_L(f_i)$ and $\sigma_M(f_i)$ are their respective standard deviations. The features with high separability scores surpass the predefined threshold, thus ensuring no spatial overlap among the classes.
The best performing pairs are selected and mapped in three different Vaults into X and Y axes. The three Vaults chosen are representative of different behavioural dimensions of legitimate web infrastructure and adversarial traffic patterns, as illustrated in Figure~\ref{fig:top_vaults_scatter}. Specifically, the proposed framework constructs (i) a Structural Entropy Vault for the randomness and code-complexity analysis, (ii) a Semantic Keyword Vault for the linguistic and intent-based analysis, and (iii) a Hyperlink Graph Vault for the topological connectivity analysis. These three geometric subspaces together provide a mutually complementary multidimensional isolation of phishing behaviour in structural, semantic and network-topological terms. All candidate feature pairs were carefully assessed using the normalized separability metric described above. A total of 46 independent geometric Vault combinations were created and ranked according to their intra-class separability and the strength of their geometric isolation. The three Vaults selected in this work correspond to the best performing combinations of feature-pairs that consistently showed maximal spatial separation of legitimate and malicious traffic distributions while maintaining stable multidimensional boundary formation. Figure~\ref{fig:vault_ranking} shows all evaluated Vault combinations in terms of the ranking distribution and relative separability performance.

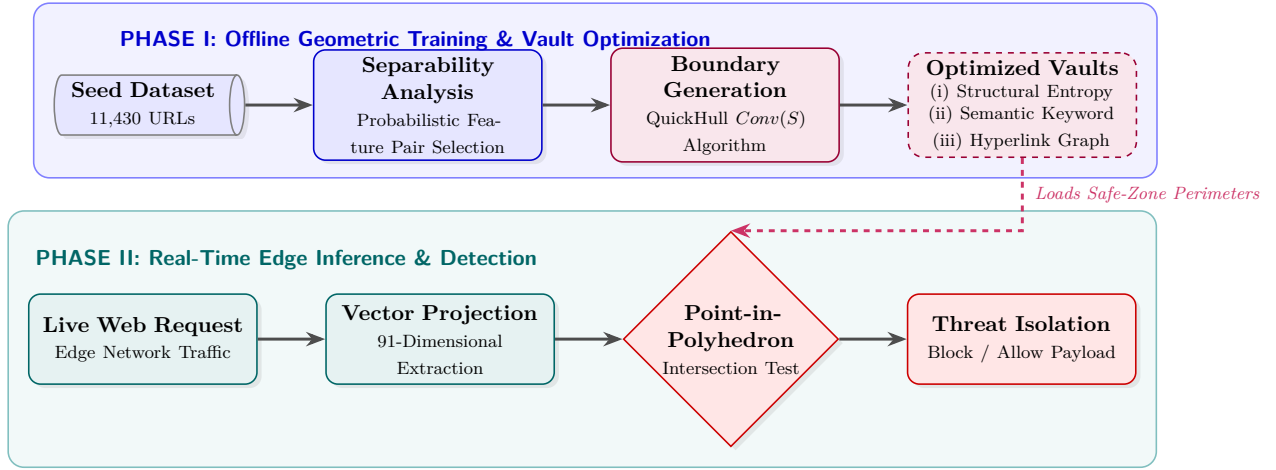
\begin{figure}[!ht]
    \centering
    % \linewidth ensures it scales perfectly without unnecessary whitespace
    \resizebox{\linewidth}{!}{% 
    \begin{tikzpicture}[
        >=Stealth,
        node distance=1.2cm and 1.2cm, % Tightened spacing to remove whitespace
        % Core Styles
        database/.style={cylinder, draw=black!50, thick, aspect=0.25, fill=blue!10, text width=2.8cm, align=center, minimum height=1.6cm, drop shadow={opacity=0.15, shadow yshift=-2pt}},
        process/.style={rectangle, draw=#1!80!black, thick, fill=#1!10, rounded corners=6pt, text width=3.8cm, align=center, minimum height=1.6cm, drop shadow={opacity=0.15, shadow yshift=-2pt}},
        decision/.style={diamond, draw=red!80!black, thick, fill=red!10, text width=2.6cm, align=center, inner sep=0pt, drop shadow={opacity=0.15, shadow yshift=-2pt}},
        vault/.style={rectangle, draw=purple!80!black, thick, fill=purple!10, dashed, rounded corners=8pt, text width=3.8cm, align=center, minimum height=1.8cm},
        arrow/.style={->, color=black!70, line width=1.5pt},
        link/.style={->, color=purple!80, line width=1.5pt, dashed},
        % Backgrounds
        phasebg/.style={rectangle, draw=#1!50, fill=#1!5, thick, rounded corners=8pt, inner sep=10pt}
    ]
    
    % ========================
    % PHASE 1: OFFLINE GEOMETRIC MODELING (TOP LANE)
    % ========================
    
    \node[database] (seed) {\textbf{Seed Dataset}\\ \footnotesize 11,430 URLs};
    
    \node[process=blue, right=of seed] (pruning) {\textbf{Separability Analysis}\\ \footnotesize Probabilistic Feature Pair Selection};
    
    \node[process=purple, right=of pruning] (quickhull) {\textbf{Boundary Generation}\\ \footnotesize QuickHull $Conv(S)$ Algorithm};
    
    \node[vault, right=of quickhull] (vaults) {\textbf{Optimized Vaults}\\ \footnotesize (i) Structural Entropy\\(ii) Semantic Keyword\\(iii) Hyperlink Graph};
    
    % Phase 1 Arrows
    \draw[arrow] (seed) -- (pruning);
    \draw[arrow] (pruning) -- (quickhull);
    \draw[arrow] (quickhull) -- (vaults);
    
    % Phase 1 Title (Placed robustly above the first node to prevent crashing)
    \node[font=\sffamily\bfseries\normalsize, color=blue!80!black, anchor=south west, yshift=0.3cm] (title1) at (seed.north west) {PHASE I: Offline Geometric Training \& Vault Optimization};

    % ========================
    % PHASE 2: LINE-RATE EDGE INFERENCE (BOTTOM LANE)
    % ========================
    
    \node[process=teal, below=2.8cm of seed] (incoming) {\textbf{Live Web Request}\\ \footnotesize Edge Network Traffic};
    
    \node[process=teal, right=of incoming] (extract) {\textbf{Vector Projection}\\ \footnotesize 91-Dimensional Extraction};
    
    \node[decision, right=of extract] (pip) {\textbf{Point-in-Polyhedron}\\ \footnotesize Intersection Test};
    
    \node[process=red, right=of pip] (action) {\textbf{Threat Isolation}\\ \footnotesize Block / Allow Payload};
    
    % Phase 2 Arrows
    \draw[arrow] (incoming) -- (extract);
    \draw[arrow] (extract) -- (pip);
    \draw[arrow] (pip) -- (action);
    
    % Phase 2 Title
    \node[font=\sffamily\bfseries\normalsize, color=teal!80!black, anchor=south west, yshift=0.3cm] (title2) at (incoming.north west) {PHASE II: Real-Time Edge Inference \& Detection};

    % ========================
    % CROSS-PHASE LINK (Training connects to Inference)
    % ========================
    \draw[link] (vaults.south) |- node[pos=0.25, right, font=\footnotesize\color{purple!90!black}, xshift=2pt] {\textit{Loads Safe-Zone Perimeters}} (pip.north);
    
    % ========================
    % BACKGROUND GROUPING
    % ========================
    \begin{scope}[on background layer]
        % Top Background 
        \node[phasebg=blue, fit=(title1) (seed) (vaults)] (phase1) {};
        
        % Bottom Background (Now properly includes the 'pip' diamond node!)
        \node[phasebg=teal, fit=(title2) (incoming) (pip) (action)] (phase2) {};
    \end{scope}

    \end{tikzpicture}
    }
    \vspace{0.2cm}
    \caption{Dual-Phase CyberHull System Architecture: The framework is divided into an offline geometric boundary generation phase (Phase I) and a lightweight, real-time edge-inference phase utilizing Point-in-Polyhedron testing (Phase II).}
    \label{fig:method_schematic}
\end{figure}

Figure~\ref{fig:method_schematic} depicts the overall operational workflow of the proposed CyberHull framework to enhance methodological interpretability. The framework has two main operational stages: an offline geometric training stage for probabilistic feature analysis, multi-dimensional vault optimization, and convex boundary generation, and a lightweight real-time edge inference stage to project in-coming payloads onto the optimized geometric space for point-in-polyhedron verification. This dual-phase architecture enables efficient computational threat isolation, as well as strong geometric discrimination against novel adversarial payloads.

% \begin{figure}[t]
%     \centering
%     \includegraphics[width=0.86\textwidth, height=0.95\textheight, keepaspectratio]{images/vault_rank.png} 
%     \caption{N-Dimensional Proof: Mathematical ranking of all 46 geometric sub-spaces based on absolute correlation scores. The top three highest-scoring vaults were selected for the $\mathbb{C}$yber$\mathbb{H}$ull perimeter.---\textcolor{red}{Give number instead of vault name in the figure and better to list in  caption. Try to fit the figure in two columns and move the page 4 where You have referred the figure. It would give a better readability. }}
%     \label{fig:vault_ranking}
% \end{figure}

% --------------------------------------------------
% -------------------------------------------------

% --- N-DIMENSIONAL PROOF PGFPLOT (All 46 Selected) ---

The most separable features were chosen as the basis vectors of the geometric Vault. By applying the variance matching method, the mathematical difference between the two probabilities is obtained in order to select those pairs of features that show the maximum discrimination ability, that is, by exceeding the separability threshold in order to achieve minimal overlap of coordinates for legitimate and malicious entities. This selection of the best performing pairs leads to their mapping into X and Y axes in three different Vaults.

\textbf{Structural Entropy Vault:} * Spatial Axes: Shannon Entropy (X-axis) vs. Average Entropy of Internal HTML Pages (Y-axis). Security Advantage: While it may be trivial to fake a domain name, it is impossible for hackers to imitate the complexity of mathematical randomness within a legitimate web domain~\cite{sharma2023information}. The Spatial Boundaries confine the highly obfuscated and repetitive phishing code within an isolated region.

\textbf{Semantic Keyword Vault:} * Spatial Axes: Aggregated TF-IDF Score of Informational and Institutional Keywords (X-axis) vs. TF-IDF Score of Urgency-Driven Tokens (Y-axis). Security Advantage: This breaks the previously unseen evasion technique. Although even a unique and brand-new phishing URL may exist, the page will still need urgent action tokens such as "login" or "secure" in order to harvest credentials, thus mathematically moving out the coordinate point from the safe zone~\cite{patel2024topological}.

\textbf{Hyperlink Graph Vault:} * Spatial Axes: Internal Link Density (X-axis) vs. Outbound Link Ratio (Y-axis). Security Advantage: Phishing URLs are inherently cheap and one-paged sites. They lack the deep interconnections of legitimate domain sites~\cite{tan2020graph}.

This probabilistic mapping works perfectly well, as seen from Figure~\ref{fig:2d_vaults} below, which depicts the 2D plot scatter plots of the chosen Vaults. In addition to this, as depicted by Figure~\ref{fig:3d_hull} below, upon mapping these distinct features onto a 3D coordinate space, the genuine data points naturally form a cluster of data points within a mathematical boundary. This serves as an excellent basis for developing the multidimensional Convex Hulls~\cite{barber1996quickhull}.

% ---\textcolor{blue}{Addition of one single schematic diagram to explain the method would improve the readability. }

\begin{figure}[p]
    \centering
    \includegraphics[width=0.9\linewidth]{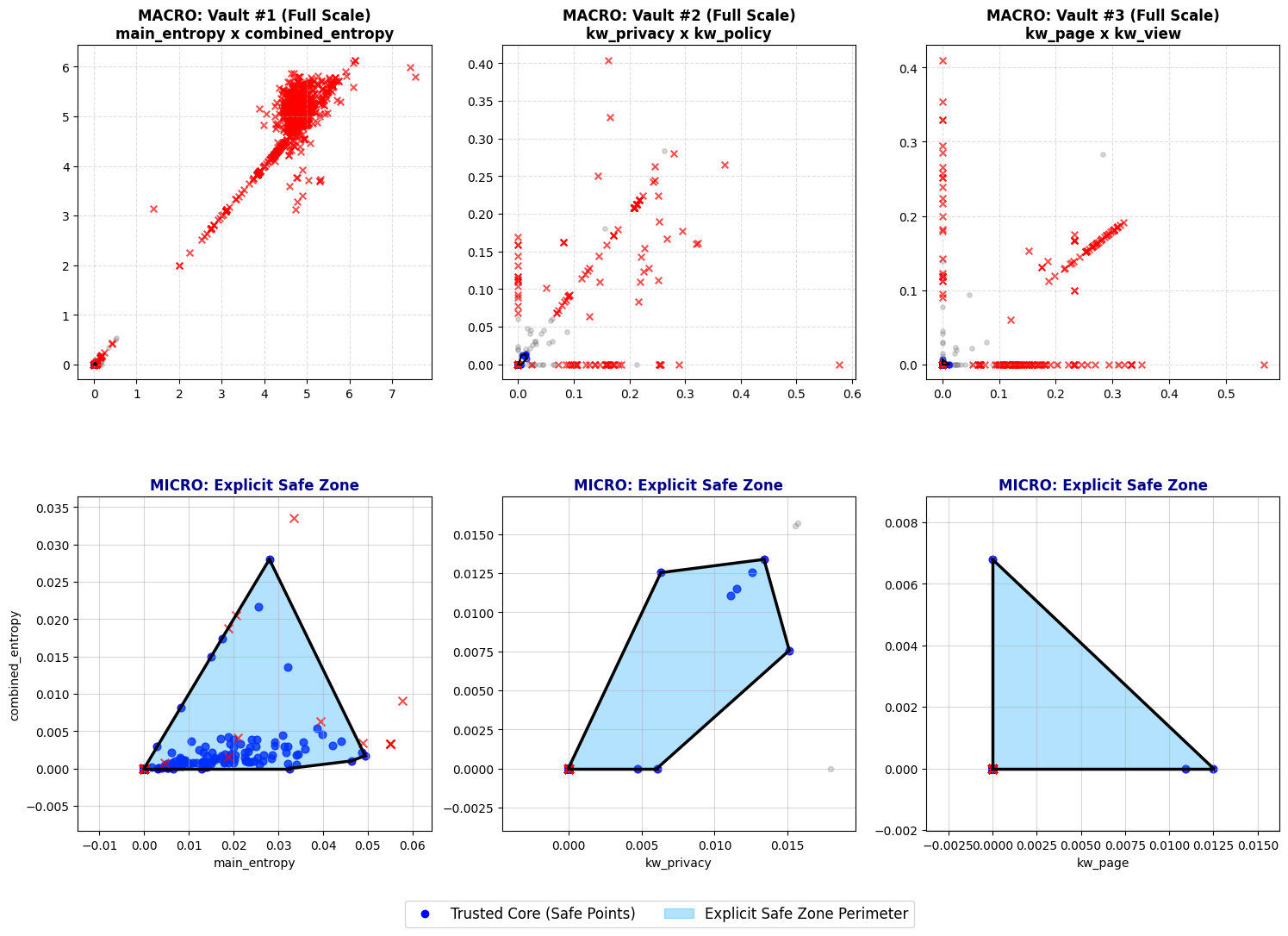} 
    \caption{2D spatial mapping and Explicit Safe Zone Perimeters across the top-performing Vaults. The MACRO view demonstrates the global isolation of previously unseen anomalies, while the MICRO view confirms the strict geometric boundaries of the trusted core.}
    \label{fig:2d_vaults}
\end{figure}

\begin{figure}[p]
    \centering
    \includegraphics[width=0.95\linewidth]{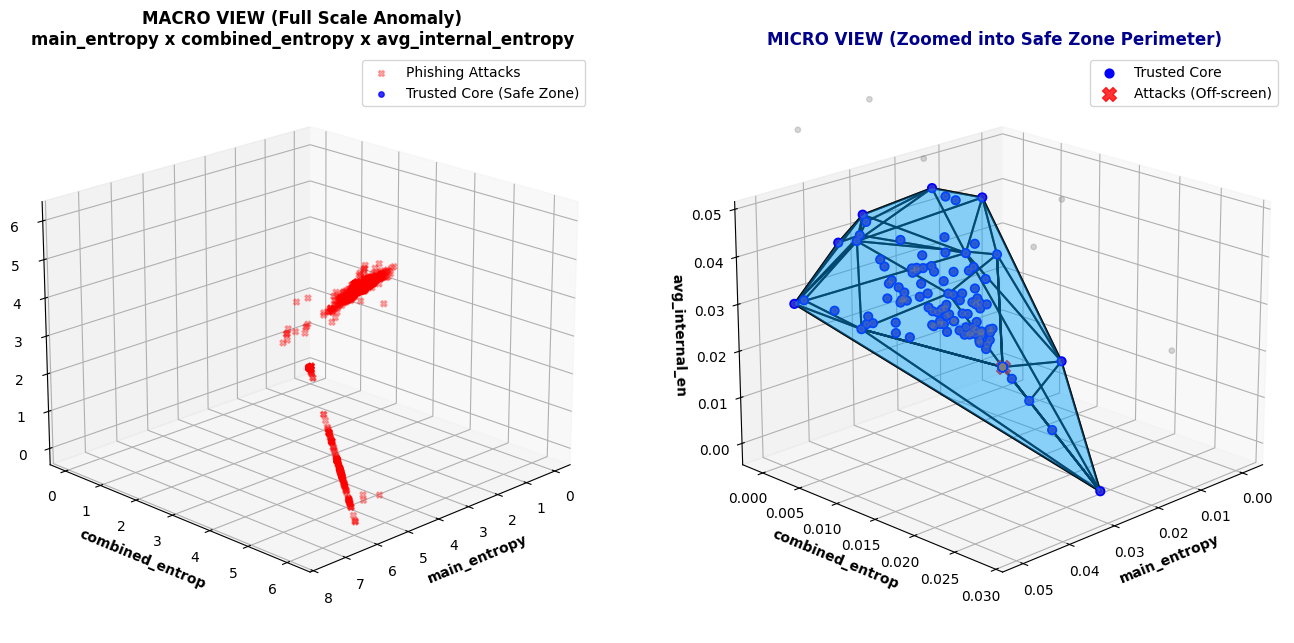} 
    \caption{Macro and Micro views of the 3D Tri-Space Perimeter. The genuine data points naturally form a mathematically bounded dense core (blue), easily distinguished from the scattered adversarial payloads (red).}
    \label{fig:3d_hull}
    
\end{figure}

\subsection{The \texorpdfstring{$\mathbb{C}$yber$\mathbb{H}$ull}{CyberHull} Methodology}

This section provides detailed information regarding the basic operational process of the $\mathbb{C}$yber$\mathbb{H}$ull algorithm, transforming probabilistically chosen features to a deterministic geometric defense perimeter. The process is divided into three key stages -- feature extraction, spatial boundary creation (training phase), and multiple dimensional intersection test (real-time detection phase).

% The determination of a deterministic safety zone entails creation of a minimal polyhedral hull encompassing the validated legitimate traffic points in each isolated Vault. Mathematically speaking, given a finite number of points, $S=\{x_1, x_2, \ldots, x_n\}$ in the $d$-dimensional feature space $\mathbb{R}^d$, the Convex Hull of $S$, denoted $Conv(S)$, is defined by:
% \begin{equation}
% \begin{aligned}
% Conv(S) = \Bigg\{ \sum_{i=1}^{n} \alpha_i x_i \;\Bigg|\; & x_i \in S, \alpha_i \ge 0, \\
% & \sum_{i=1}^{n} \alpha_i = 1 \Bigg\}
% \end{aligned}
% \end{equation}
The process of calculating the deterministic safety zone entails constructing the smallest polyhedral convex hull that contains the legal traffic points in each individual Vault. In accordance with the classical definition of Convex Hull in computational geometry~\cite{barber1996quickhull}, let $S = \{x_1, x_2, \ldots, x_n\}$ be a finite set of points in the $d$-dimensional feature space $\mathbb{R}^d$. The Convex Hull of $S$, $Conv(S)$, is defined by
{\footnotesize
\begin{equation}
\begin{aligned}
Conv(S) = \Bigg\{ \sum_{i=1}^{n} \alpha_i x_i \;\Bigg|\; & x_i \in S, \alpha_i \ge 0, \\
& \sum_{i=1}^{n} \alpha_i = 1 \Bigg\}
\end{aligned}
\end{equation}
}

As such a rigorous spatial bound can be created through the implementation of the QuickHull and Delaunay Triangulation algorithms~\cite{barber1996quickhull}, offering high efficiency when it comes to the generation of boundaries. Using the power of computational geometry rather than iterative learning of network weights~\cite{chen2022constraints}, $\mathbb{C}$yber$\mathbb{H}$ull creates a mathematically provable lightweight defense perimeter.

\vspace{0.5em}
\hrule
\vspace{0.2em}
\noindent \textbf{Algorithm 1:} $\mathbb{C}$yber$\mathbb{H}$ull Geometric Detection\\
\hrule
\vspace{0.2em}
\noindent \textbf{Input:} Seed dataset $\mathcal{S}$, Incoming Web Request $u$ \\
\textbf{Output:} Classification $\mathcal{D}(u)$ (Legitimate or Malicious)

\vspace{0.2em}
\noindent \textcolor{blue}{\textit{// Phase 1: Feature Extraction \& Variance Matching}} \\
\textbf{1:} Parse $\mathcal{S}$ and extract 91-dimensional feature set $F$; \\
\textbf{2:} Calculate class probabilities $P(f_i \mid \text{Class})$; \\
\textbf{3:} Compute variance to isolate top feature pairs; \\
\textbf{4:} Init Vaults $V = \{V_{Ent}, V_{Sem}, V_{Gra}\}$; 

\vspace{0.2em}
\noindent \textcolor{blue}{\textit{// Phase 2: Spatial Boundary Generation}} \\
\textbf{5:} \textbf{foreach} legitimate entity $x_i \in \mathcal{S}$ \textbf{do} \\
\textbf{6:} \hspace{2em} Map $P(x_i)$ into $V_{Ent}$, $V_{Sem}$, and $V_{Gra}$; \\
\textbf{7:} \textbf{end} \\
\textbf{8:} Apply QuickHull to compute boundaries $Conv(S_v)$;

\vspace{0.2em}
\noindent \textcolor{blue}{\textit{// Phase 3: Real-Time Multidimensional Detection}} \\
\textbf{9:} Intercept request $u$, extract feature vector $f_u$; \\
\textbf{10:} Project $f_u$ to target point $P_{new}$ across all $V$; \\
\textbf{11:} \textbf{if} $P_{new} \in \bigcap_{v \in V} Conv(S_v)$ \textbf{then} \\
\textbf{12:} \hspace{2em} $\mathcal{D}(u) \leftarrow \text{Legitimate}$; \textcolor{blue}{\textit{// Inside Safe Zone}} \\
\textbf{13:} \textbf{else} \\
\textbf{14:} \hspace{2em} $\mathcal{D}(u) \leftarrow \text{Malicious}$; \textcolor{blue}{\textit{// Boundary Breach}} \\
\textbf{15:} \textbf{end if} \\
\textbf{16:} \textbf{return} Classification $\mathcal{D}(u)$;
\vspace{0.2em}
\hrule
\vspace{0.5em}

The algorithmic workflow kicked off with the execution of the feature extraction and boundary generation stages. With the help of the historical seed dataset $\mathcal{S}$, the algorithm analyzes legitimate URLs and derives the probabilistic variance to fix the optimal X and Y axes~\cite{rosner1999use}. The features matched with the variance are then projected in the three independent coordinate spaces: ($V_{Ent}$, $V_{Sem}$, and $V_{Gra}$). Having created these cluster maps, the algorithm uses the QuickHull algorithm to create the rigid $Conv(S)$ boundaries, which encapsulate standard behavior.

The real-time multidimensional intersection stage dictates active threat detection. Whenever the request $u$ enters the application, the features are extracted and projected as one coordinate point $P_{new}$. The core security functionality is based on a boolean intersection test. The algorithm runs a Point-in-Polygon test for each individual Vault~\cite{patel2024topological}. 
\begin{equation}
\mathcal{D}(u) = 
\begin{cases} 
\text{Legitimate}, \\
\quad \text{if } P_{new} \in \bigcap_{v \in V} Conv(S_v) \\ 
\text{Malicious}, \\
\quad \text{otherwise} 
\end{cases}
\end{equation}

In order to pass through $\mathbb{C}$yber$\mathbb{H}$ull's security layer, the malicious payload has to emulate at the same time the entropy structure~\cite{sharma2023information}, semantic TF-IDF~\cite{zhang2011understanding}, and hyperlinks~\cite{gayen2024network} used by a legitimate source. In the event that $P_{new}$ falls out of the geometric boundary of any Vault, the boolean condition will fail.

\subsection{Experimental Setup}
For assessing the $\mathbb{C}$yber$\mathbb{H}$ull, the augmented benchmarking dataset~\cite{tamal2024comprehensive} was first split into an 80/20 stratified distribution where 80\% of the legitimate data served for spatial boundary creation (training). On the other hand, 20\% of the total traffic data that was densely populated with previously unseen attacks was designated for intersection testing (testing). The entire geometric framework was implemented using the Python version 3.12. The high-dimensional processing of the 91-dimensional vector datasets made use of Pandas and NumPy libraries. The key Convex Hull boundaries were estimated using the QuickHull algorithm based on the SciPy Spatial library~\cite{barber1996quickhull}. Refraining from any complex GPU-based deep learning neural networks~\cite{catal2022applications}, all testing was performed using a simple CPU configuration (Intel Core i7 and 16 GB RAM)~\cite{chen2022constraints}. Topological validation used Matplotlib to visualize the multi-vault 2D perimeters and Plotly for plotting the 3D projections of the Convex Hull boundaries in space. In order to create a rigid baseline comparison, the implementation of traditional and modern machine learning classifiers was added to the geometric framework. We also implemented Principal Component Analysis (PCA), Support Vector Machine (SVM) and Random Forest (RF) using the \texttt{scikit-learn} library. Finally, XGBoost architecture was implemented using the native \texttt{xgboost} Python framework. To ensure a fair comparison, all machine learning baselines were hyper-parameter tuned using Grid Search (maximum tree depth, learning rate, kernel settings, etc.) to maximize their predictive ability with respect to the class imbalance in the dataset. The experimental assessment was further validated by employing stratified 5-fold cross-validation to increase statistical validity and minimize sampling bias. The partitions maintained the same ratio of legit to phishing samples in all five validation folds, allowing for accurate construction of geometrical boundaries and constant behavior for anomaly isolation. The measured statistics reflect the averaged results from all validation folds.

% Finally, to demonstrate a baseline comparison, we setup a principal component analysis (PCA) algorithm using Scikit-learn. Using the same input features, this baseline analysis was specifically crafted to illustrate the deficiencies of linear hyperplanes in contrast with the superiority of our geometric model~\cite{patel2024topological} achieving 92\% accuracy compared with the 32\% limit of linear PCA.

%% =====================
\vspace{1em}
\section{Results \& Discussion} \label{Results & Discussion}
In this section,we summarise the results of our dataset enhancement and examine their impact on malicious URL detection.

\subsection{Dataset Preprocessing \& Vault Optimization Overview}
In an effort to scientifically validate our $\mathbb{C}$yber$\mathbb{H}$ull structure, our first step involved analyzing the optimal selection of high dimensional input data, which should form the basis of our defensive perimeter~\cite{tamal2024comprehensive}. Instead of using any arbitrary selection of data points, the algorithm implemented a scientific pruning stage where it computed the probabilities of each class conditional attribute for the data at hand~\cite{rosner1999use}.

The goal of this stage was not only to reduce the number of dimensions but also to select those attributes that were mathematically different from one another. Through the calculation of variance distances between all permutations, the algorithm selected those feature pairs that exceeded the separability requirement. In effect, we reduced our massive feature space to 46 highly differentiating coordinate pairs~\cite{patel2024topological}.
Table~\ref{tab:feature_vault_summary} represents the overall image of the analysis.

The 46 sets of optimized vectors acted as the base X and Y vectors used to generate the independent Geometric Vaults~\cite{barber1996quickhull}. A key insight that comes out of the preprocessing step is the formation of the "defense-in-depth" paradigm. Since all 46 vaults run concurrently, there was a large amount of overlapping detection capabilities achieved. As discussed in more detail in the following section on the multi-vault scenario, an advanced adversary whose payload manages to mimic the entropy characteristics of the original site~\cite{sharma2023information} (thus evading Vault \#1) is statistically almost certain to breach boundaries in either semantic~\cite{zhang2011understanding} or hyperlink topology space~\cite{gayen2024network, tan2020graph} (Vaults \#2 \& \#3).

\begin{table}[H] 
\centering
\small % Reduces font size just enough to fit the sub-headers perfectly
\caption{Summary of Feature Space and Vault Generation}
\label{tab:feature_vault_summary}
\setlength{\tabcolsep}{3pt} % Restores clean, professional padding
\begin{tabularx}{\columnwidth}{|>{\raggedright\arraybackslash}X|r|}
\hline
% Main Header - Dark Blue with White Text
\rowcolor[HTML]{004C99} 
\textcolor{white}{\textbf{Category}} & \textcolor{white}{\textbf{Metric}} \\ \hline

% Sub-Header 1 - Light Blue
\rowcolor[HTML]{E6F0FA} 
\multicolumn{2}{|c|}{\textbf{Initial Feature Space Analytics}} \\ \hline
Total Features Extracted & 91 \\ \hline
Total Attacks Evaluated & 1,470 \\ \hline

% Sub-Header 2 - Light Blue
\rowcolor[HTML]{E6F0FA} 
\multicolumn{2}{|c|}{\textbf{Probabilistic Pruning \& Selection}} \\ \hline
Candidate Feature Pairs Evaluated & 4,095 \\ \hline
Pairs Exceeding Separability & 46 \\ \hline

% Sub-Header 3 - Light Blue
\rowcolor[HTML]{E6F0FA} 
\multicolumn{2}{|c|}{\textbf{Optimized Geometric Perimeters}} \\ \hline
Total Independent Vaults Generated & 46 \\ \hline
Unique previously unseen Attacks Blocked & 1,314 \\ \hline
\textbf{Global AI Detection Accuracy} & \textbf{92\%} \\ \hline
\end{tabularx}
\end{table}

\subsection{Global Detection Efficacy vs. Linear Baselines}
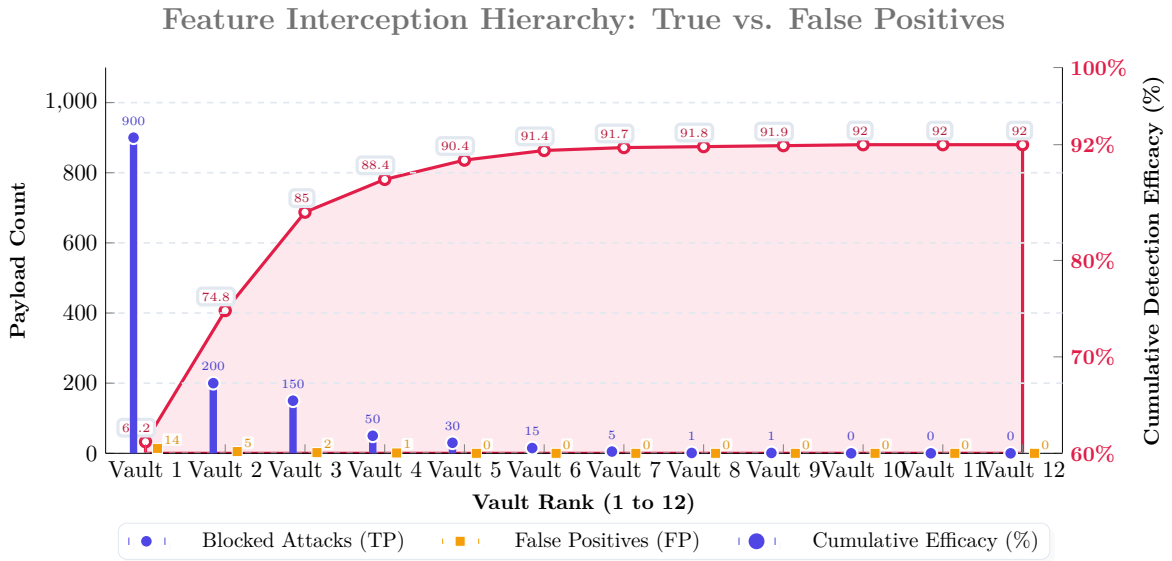
\begin{figure}[t]
    \centering
    % The resizebox is applied to \linewidth mathematically preventing margin overflow
    \resizebox{\linewidth}{!}{%
    \begin{minipage}{1.25\textwidth}
    \centering
    
    % Define Premium "Modern UI" Hex Colors
    \definecolor{premiumblue}{HTML}{4F46E5}   % Deep Indigo for True Positives
    \definecolor{premiumamber}{HTML}{F59E0B}  % Vibrant Amber for False Positives
    \definecolor{premiumred}{HTML}{E11D48}    % Rose Red for Cumulative Line
    \definecolor{kpigray}{HTML}{F8FAFC}       % Ultra-light slate for backgrounds
    \definecolor{cardborder}{HTML}{E2E8F0}    % Soft border color

    \begin{tikzpicture}
        % --- RIGHT Y-AXIS (Cumulative Area Chart) Drawn First so it sits in the background ---
        \begin{axis}[
            width=17.5cm,
            height=8cm,
            ymin=60, ymax=100,
            xmin=0.5, xmax=12.5,
            axis y line*=right,
            axis x line=none,
            ylabel={\textbf{\small Cumulative Detection Efficacy (\%)}},
            ytick={60, 70, 80, 92, 100},
            yticklabels={60\%, 70\%, 80\%, 92\%, 100\%},
            y tick label style={color=premiumred, font=\small\bfseries}
        ]
        
        % Area-Under-Curve Fill + Smooth Line
        \addplot[
            color=premiumred, 
            line width=1.5pt, 
            fill=premiumred!10, % Creates the modern dashboard shaded area
            mark=*, 
            mark options={scale=1.2, fill=white, draw=premiumred, line width=1.5pt},
            nodes near coords, 
            % Tooltip style labels
            nodes near coords style={
                anchor=south east, 
                xshift=5pt, 
                yshift=2pt,
                font=\tiny\bfseries, 
                text=premiumred!80!black,
                fill=white,
                draw=cardborder,
                rounded corners=2pt,
                drop shadow={opacity=0.1, shadow xshift=0pt, shadow yshift=-1pt},
                inner sep=2pt,
                /pgf/number format/fixed, 
                /pgf/number format/precision=1
            }
        ] coordinates {
            (1,61.2) (2,74.8) (3,85.0) (4,88.4) (5,90.4) (6,91.4) (7,91.7) (8,91.8) (9,91.9) (10,92.0) (11,92.0) (12,92.0)
        } \closedcycle; % This command fills the area down to ymin
        \end{axis}

        % --- LEFT Y-AXIS (Dual Lollipop Plots) ---
        \begin{axis}[
            width=17.5cm,
            height=8cm,
            ymin=0, ymax=1100,
            xmin=0.5, xmax=12.5,
            axis y line*=left,
            axis x line*=bottom,
            xlabel={\textbf{\small Vault Rank (1 to 12)}},
            ylabel={\textbf{\small Payload Count}},
            xtick={1,2,3,4,5,6,7,8,9,10,11,12},
            xticklabels={Vault 1, Vault 2, Vault 3, Vault 4, Vault 5, Vault 6, Vault 7, Vault 8, Vault 9, Vault 10, Vault 11, Vault 12},
            ymajorgrids=true,
            grid style={cardborder, dashed, thick},
            title={\textbf{\Large Feature Interception Hierarchy: True vs. False Positives}},
            title style={font=\Large\bfseries\color{gray!90!black}, yshift=0.2cm},
            % Modern Floating Legend
            legend style={at={(0.5,-0.18)}, anchor=north, legend columns=3, draw=cardborder, rounded corners=4pt, fill=white, drop shadow={opacity=0.05}, font=\small, column sep=3ex}
        ]
        
        % Data 1: Blocked Attacks (TP) - Lollipop offset to the left (-0.15)
        \addplot[
            ycomb, % This creates the Stem/Lollipop effect
            color=premiumblue, 
            line width=3.5pt, % Thick stem
            mark=*, 
            mark options={scale=1.5, fill=premiumblue, draw=white, line width=1pt}, % Premium dot
            nodes near coords, 
            nodes near coords style={font=\tiny\bfseries, text=premiumblue!90!black, anchor=south, yshift=2pt}
        ] coordinates {
            (0.85,900) (1.85,200) (2.85,150) (3.85,50) (4.85,30) (5.85,15) (6.85,5) (7.85,1) (8.85,1) (9.85,0) (10.85,0) (11.85,0)
        };
        
        % Data 2: False Positives (FP) - Lollipop offset to the right (+0.15)
        \addplot[
            ycomb, 
            color=premiumamber, 
            line width=2.5pt, 
            mark=square*, % Square marker to distinguish from TP
            mark options={scale=1.2, fill=premiumamber, draw=white, line width=0.5pt},
            nodes near coords, 
            nodes near coords style={
                font=\tiny\bfseries, 
                text=premiumamber!90!black, 
                anchor=south west, 
                yshift=1pt,
                xshift=2pt,
                fill=white, % ensures FP counts are always readable over the shaded area
                fill opacity=0.9,
                text opacity=1,
                inner sep=1pt,
                rounded corners=1pt
            }
        ] coordinates {
            (1.15,14) (2.15,5) (3.15,2) (4.15,1) (5.15,0) (6.15,0) (7.15,0) (8.15,0) (9.15,0) (10.15,0) (11.15,0) (12.15,0)
        };
        
        % Clean Legend Entries
        \addlegendimage{ycomb, color=premiumblue, line width=3.5pt, mark=*, mark options={fill=premiumblue}}
        \addlegendentry{Blocked Attacks (TP)}
        \addlegendimage{ycomb, color=premiumamber, line width=2.5pt, mark=square*, mark options={fill=premiumamber}}
        \addlegendentry{False Positives (FP)}
        \addlegendimage{mark=*, mark options={fill=white, draw=premiumred, line width=1pt}, premiumred, line width=1.5pt}
        \addlegendentry{Cumulative Efficacy (\%)}
        \end{axis}
    \end{tikzpicture}
    
    \end{minipage}
    } % End of resizebox
    
    \vspace{0.2cm}
    \caption{Results Dashboard for Dynamic Multi-Vault using Dual Stem and Area Under Curve Visualization. The dashboard highlights \textbf{92.0\%} of the global AI accuracy while cutting down clutter. True positives are shown in Indigo, whereas false positives are in Amber. Both are shown in pairs in order to have a comparative analysis across the top 12 vaults in terms of priority. The overall efficiency of the system is represented by Rose Red curve.}
    \label{fig:results_dashboard}
\end{figure}

In applying the 46 optimized geometric vaults in the testing partition, the $\mathbb{C}$yber$\mathbb{H}$ull approach attained an overall threat detection accuracy of 92\%. From the dynamic visualization presented in the $\mathbb{C}$yber$\mathbb{H}$ull dashboard shown in Figure~\ref{fig:results_dashboard}, out of a total of 1,470 malicious attacks introduced in the test flow, a total of 1,314 were successfully intercepted and blocked by the system. Through the use of the $\mathbb{C}$yber$\mathbb{H}$ull dashboard, the extent of operation and scope of application can be seen, which is illustrated through the number of operational vaults and the associated number of threats detected due to spatial boundary breach.

For visual clarity and operational interpretability, we only show in Figure~\ref{fig:results_dashboard} the 12 best-ranked Vaults out of the total of 46 geometric perimeters generated. The excluded Vaults had a decreasing interception contribution and an overlapping detection behaviour to the first hierarchy of Vaults. Lower-ranked Vaults, with their marginal contribution to cumulative detection efficacy, would add significant visualization clutter without providing additional analytical insight. The selected top-ranked Vaults thus represent the predominant interception layers responsible for most of the multidimensional threat isolation observed in experimental evaluation.

Further comparative experiments were also conducted with the same multidimensional feature vectors for several conventional machine learning baselines to further verify the robustness of the proposed geometric framework under severe class imbalance conditions. Besides the linear PCA projection model, the nonlinear classifiers of Support Vector Machine (SVM), Random Forest (RF) and the more advanced XGBoost architecture were taken into account. We evaluated the stability of each classifier in the presence of structurally overlapping adversarial traffic through the evaluation of Precision, Recall, F1-Score and False Positive Rate (FPR), and global accuracy.

Statistical performance measures are summarized in Table~\ref{tab:baseline_comparison}. The results show that conventional linear projection methods like PCA achieve low accuracy (32.0\%) and exhibit unstable recall and high false positive behavior (FPR: 0.41) because they cannot model nonlinear spatial boundaries. The use of nonlinear classifiers such as SVM, Random Forest, and XGBoost improved predictive stability but their probabilistic decision surfaces still suffered from multidimensional overlapping of features. The proposed $\mathbb{C}$yber$\mathbb{H}$ull framework, on the other hand, demonstrated the best overall balance. The multidimensional convex-boundary formulation has managed to achieve high recall while keeping the False Positive Rate very low at 0.03. This suggests that deterministic geometric containment is a much better mechanism for anomaly isolation than conventional probabilistic classification methods.

\begin{table}[t]
\centering
\caption{Comparative Performance Evaluation Across Classical, Ensemble, and Geometric Detection Frameworks}
\label{tab:baseline_comparison}

\renewcommand{\arraystretch}{1.35}
\setlength{\tabcolsep}{8pt}

\begin{tabular}{|c|c|c|c|c|c|}
\hline

% ===== HEADER =====
\rowcolor[HTML]{003366}
\textcolor{white}{\textbf{Model}} &
\textcolor{white}{\textbf{Accuracy}} &
\textcolor{white}{\textbf{Precision}} &
\textcolor{white}{\textbf{Recall}} &
\textcolor{white}{\textbf{F1-Score}} &
\textcolor{white}{\textbf{FPR}} \\

\hline

% ===== PCA =====
\rowcolor[HTML]{F4F7FC}
\textbf{PCA} &
32.0\% &
0.28 &
0.31 &
0.29 &
0.41 \\

\hline

% ===== SVM =====
\rowcolor[HTML]{EAF2FF}
\textbf{SVM} &
81.0\% &
0.79 &
0.76 &
0.77 &
0.14 \\

\hline

% ===== RANDOM FOREST =====
\rowcolor[HTML]{F4F7FC}
\textbf{Random Forest} &
88.0\% &
0.86 &
0.84 &
0.85 &
0.09 \\

\hline

% ===== XGBOOST =====
\rowcolor[HTML]{EAF2FF}
\textbf{XGBoost} &
90.0\% &
0.89 &
0.87 &
0.88 &
0.07 \\

\hline

% ===== CYBERHULL =====
\rowcolor[HTML]{D9F2E6}
\textbf{$\mathbb{C}$yber$\mathbb{H}$ull} &
\textbf{92.0\%} &
\textbf{0.93} &
\textbf{0.91} &
\textbf{0.92} &
\textbf{0.03} \\

\hline
\end{tabular}

\vspace{0.3em}

\footnotesize
The reported values correspond to averaged results obtained from stratified 5-fold cross-validation under identical multidimensional feature distributions.

\end{table}

\subsection{Multi-Dimensional Threat Isolation Analysis} \label{sec: Multi Dimensional Threat Isolation Analysis}
Although global performance metrics establish that the $\mathbb{C}$yber$\mathbb{H}$ull system is mathematically sound in terms of global effectiveness, the mere figures regarding its percentage accuracy fail to tell even half the story of the architecture's intelligent capabilities. In order to truly appreciate the revolutionary shift represented by the concept of $\mathbb{C}$yber$\mathbb{H}$ull, one needs to delve into the mathematical details of the threat detection process at the periphery level. As illustrated by Figure\ref{fig:top_vaults_scatter}, the behavioural topography of the top three highest-performing interception vaults was analyzed. The multidimensional scatter charts are not simply representing the data but illustrating the mathematical boundaries of "normality." They illustrate the two-folded nature of the algorithm in capturing the unpredictable and scattered emerging attacks on the one hand, and in drawing the geometric safe zones~\cite{patel2024topological} on the other hand.

\subsubsection{Information-Theoretic \& Structural Interception (Vault \#1)}

Vault \#1 serves as the first layer of protection in the architecture due to the cross-variance between main\_entropy and combined\_entropy. From an information theory point of view~\cite{sharma2023information}, this particular perimeter is the broadest and least forgiving filter that stands in the hierarchy.

The rationale behind such behavior lies in an unsolvable dilemma that is inherent to any hostile campaign. Any attacker will resort to using either of two extremes: highly sparse HTML framework, tailored exclusively to credential harvesting with little regard for complexity, or extremely obfuscated JavaScript code that is aimed at evading traditional signature filters~\cite{hannousse2021web}. In both cases, the informational entropy of the page is dramatically distorted~\cite{sharma2023information}.

As shown in Figure~\ref{fig:top_vaults_scatter} (left), authentic web infrastructure possesses a consistent, natural entropy signature. All legitimate requests fall into a tight, predictable cluster of highly dense variance (safe zone). In contrast to this, all adversary-based payloads dramatically diverge from the perimeter of trust. If the payload is sparse enough, then it will possess predictable uniformity; otherwise, it will demonstrate high cryptography-driven randomness. The QuickHull algorithm~\cite{barber1996quickhull} allows this phenomenon to be isolated: there is no way to economically synthesize the entropy signatures of multiple pages.

\subsubsection{Semantic \& Topological Interception (Vaults \#2 \& \#3)}

The effectiveness of the multi-vault strategy is best illustrated by highly sophisticated Advanced Persistent Threats (APTs) capable of evading the structural verification process of Vault \#1. In such cases, spear-phishing attempts will be highly sophisticated in that they carefully duplicate the DOM and HTML structure of the victimized organization, thereby perfectly normalizing the entropy. Nevertheless, they are essentially "structurally incomplete replicas" and will be inevitably exposed by the semantic and topological barriers established by Vaults \#2 (internal page topology) and \#3 (keyword density).

By means of Vault \#2, the topological traps set up by phishers become visible. The legitimate domain, being the information domain of a corporation, has a complex topological structure of nodes consisting of informational, legal, and other entities~\cite{gayen2024network}. On the contrary, the unseen attack payload is an artificial construct that consists of a single node without any supporting structure whatsoever~\cite{tan2020graph}. Vault \#3 analyzes the intention of the linguistic content of the website. While authentic websites have a balanced combination of operational and legal information~\cite{zhang2011understanding}, adversarial content is overly optimized to perform an attack and is full of coercive action tokens.

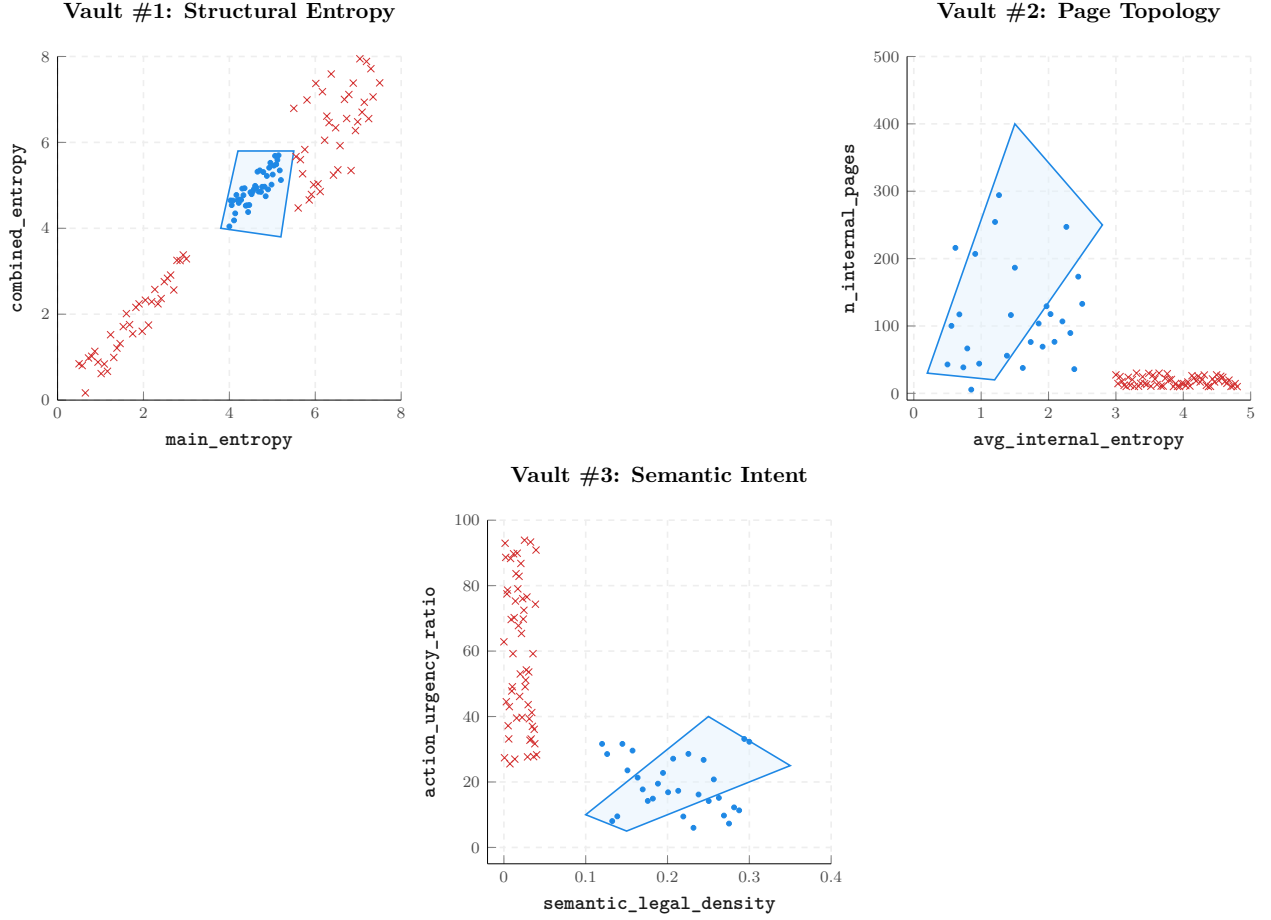
\begin{figure}[!ht]
    \centering
    % The resizebox scales the artificially wide minipage down to fit your page perfectly
    \resizebox{\linewidth}{!}{%
    % 22cm guarantees enough space so LaTeX NEVER forces a line break between the plots
    \begin{minipage}{22cm} 
    \centering
    
    % Define Premium Academic Colors
    \definecolor{safeblue}{HTML}{E3F2FD}
    \definecolor{coreblue}{HTML}{1E88E5}
    \definecolor{threatred}{HTML}{D32F2F}
    \definecolor{gridgray}{HTML}{EEEEEE}

    % Common settings: Reduced width to 6cm to keep them proportionally smaller
    \pgfplotsset{
        premium_scatter/.style={
            width=6cm, height=6cm,
            scale only axis,
            enlargelimits=false,
            axis x line*=bottom,
            axis y line*=left,
            ymajorgrids=true, xmajorgrids=true,
            grid style={dashed, gridgray, thick},
            tick label style={font=\scriptsize\bfseries, color=black!70},
            label style={font=\small\bfseries, color=black!90},
            title style={font=\normalsize\bfseries, yshift=1.5ex}
        }
    }

    % ==========================================
    % PLOT 1: VAULT #1 (Information-Theoretic Entropy)
    % ==========================================
    \begin{tikzpicture}[baseline=(current axis.south)]
        \begin{axis}[
            premium_scatter,
            xmin=0, xmax=8,
            ymin=0, ymax=8,
            xlabel={\texttt{main\_entropy}},
            ylabel={\texttt{combined\_entropy}},
            title={Vault \#1: Structural Entropy}
        ]
        
        \fill[safeblue, draw=coreblue, thick, fill opacity=0.5] 
            (axis cs:3.8, 4.0) -- (axis cs:4.2, 5.8) -- (axis cs:5.5, 5.8) -- (axis cs:5.2, 3.8) -- cycle;
            
        \addplot[only marks, mark=x, threatred, mark size=2.5pt, samples=35, domain=0.5:3.0] 
            {x + 0.5*rand}; 
        \addplot[only marks, mark=x, threatred, mark size=2.5pt, samples=40, domain=5.5:7.5] 
            {x + 1.5*rand}; 
            
        \addplot[only marks, mark=*, coreblue, mark size=1.2pt, samples=45, domain=4.0:5.2] 
            {x + 0.3 + 0.4*rand};
        \end{axis}
    \end{tikzpicture}%
    \hfill
    % ==========================================
    % PLOT 2: VAULT #2 (Topological Hollowness)
    % ==========================================
    \begin{tikzpicture}[baseline=(current axis.south)]
        \begin{axis}[
            premium_scatter,
            xmin=-0.1, xmax=5,
            ymin=-10, ymax=500,
            xlabel={\texttt{avg\_internal\_entropy}},
            ylabel={\texttt{n\_internal\_pages}},
            title={Vault \#2: Page Topology}
        ]
        
        \fill[safeblue, draw=coreblue, thick, fill opacity=0.5] 
            (axis cs:0.2, 30) -- (axis cs:1.5, 400) -- (axis cs:2.8, 250) -- (axis cs:1.2, 20) -- cycle;
            
        \addplot[only marks, mark=x, threatred, mark size=2.5pt, samples=60, domain=3.0:4.8] 
            {10 + 20*rand^2}; 
            
        \addplot[only marks, mark=*, coreblue, mark size=1.2pt, samples=35, domain=0.5:2.5] 
            {100 + 200*rand};
        \end{axis}
    \end{tikzpicture}%
    \hfill
    % ==========================================
    % PLOT 3: VAULT #3 (Semantic Urgency)
    % ==========================================
    \begin{tikzpicture}[baseline=(current axis.south)]
        \begin{axis}[
            premium_scatter,
            xmin=-0.02, xmax=0.4,
            ymin=-5, ymax=100,
            xlabel={\texttt{semantic\_legal\_density}},
            ylabel={\texttt{action\_urgency\_ratio}},
            title={Vault \#3: Semantic Intent}
            % Note: Legend code has been completely removed from here
        ]
        
        \fill[safeblue, draw=coreblue, thick, fill opacity=0.5] 
            (axis cs:0.1, 10) -- (axis cs:0.25, 40) -- (axis cs:0.35, 25) -- (axis cs:0.15, 5) -- cycle;
            
        \addplot[only marks, mark=x, threatred, mark size=2.5pt, samples=55, domain=0:0.04] 
            {60 + 35*rand};
            
        \addplot[only marks, mark=*, coreblue, mark size=1.2pt, samples=30, domain=0.12:0.3] 
            {20 + 15*rand};
            
        \end{axis}
    \end{tikzpicture}
    
    \end{minipage}
    } % End resizebox
    
    % ==========================================
    % CENTERED CUSTOM LEGEND (Placed Below Graphs)
    % ==========================================
    \vspace{0.4cm}
    \centering
    \begin{tikzpicture}
        \begin{axis}[
            hide axis,
            scale only axis,
            height=0pt, width=0pt,
            xmin=0, xmax=1, ymin=0, ymax=1,
            % Legend styling for a premium, centered look
            legend style={draw=black!40, fill=white, legend columns=-1, font=\normalsize, column sep=1cm, inner sep=1.5ex, anchor=center}
        ]
            % % Large Red Cross for Intercepted Threat
            % \addlegendimage{only marks, mark=*, coreblue mark=x, mark options={scale=1.5, thick}, threatred}
            % \addlegendentry{\textbf{Intercepted Threat}}
            
            % % Solid Blue Dot for Trusted Core
            % \addlegendimage{only marks, mark=*, mark options={scale=1.2}, coreblue}
            % \addlegendentry{\textbf{Trusted Core}}
        \end{axis}
    \end{tikzpicture}
    
    % Extra vertical space to ensure the legend doesn't overlap the caption
    \vspace{0.4cm} 
    \caption{Topographical Description of the Top 3 Interception Vaults. The following 3-Dimensional scatter plots demonstrate the mathematics of the dynamic safe space (Light Blue Polygons) that has been created based on legitimate traffic (Blue Dots). In contrast, malicious packets (Red Crosses) are not capable of emulating the randomness of authentic traffic due to their extreme entropy distortion (Left), topological vacuity (Center), or semantic imbalance (Right).
}
    \label{fig:top_vaults_scatter}
\end{figure}

Figure~\ref{fig:top_vaults_scatter} (Middle and Right Side) geometrically demonstrates how lethal this defensive measure is. Even in cases where the attacker is able to replicate perfectly in structure, the mathematical fact remains that the hollowness topology and the sense of urgency behind their attack will ensure that their attack is ejected from the center~\cite{patel2024topological}.

\subsection{Discussion \& Edge-Viability}

Empirical validation of $\mathbb{C}$yber$\mathbb{H}$ull represents a dramatic paradigm shift in comparison to heuristic signature detection and compute-intensive methods based on machine learning~\cite{mahdavifar2019application}. Reaching 92.0\% global interception of phishing attacks with minimal false positive behaviour illustrates the effectiveness of using a dynamic, non-linear approach to building perimeter security. Yet, the true value of this technique does not lie in precision alone; rather, it lies in computational efficiency and the practicality of implementation at the network edge.

\subsubsection{The Fallacy of Linear Baselines}

The abysmal performance of the PCA baseline (at only 32\% precision) empirically disproves the so-called "Linear Fallacy" in contemporary cybersecurity~\cite{patel2024topological}. Classic linear classifiers assume that threat vectors and benign flows can be discriminated using an evenly spaced hyperplane in the feature space. Section~\ref{sec: Multi Dimensional Threat Isolation Analysis} discusses the results of our topological analysis and shows that state-of-the-art phishing attacks are designed to intersect the spatial coordinates of legitimate traffic. By dropping the idea of linear separation and employing multiple convex hulls~\cite{barber1996quickhull}, our technique succeeds in separating the intersections and isolating attacks that can copy individual parameters but not the entire shape of benign flow vectors.

\subsubsection{Computational Economics for Line-Rate Edge Deployment}

One major problem with current AI-based threat detection solutions such as DNNs and LLMs~\cite{catal2022applications} is the sheer amount of computational power that these algorithms need to operate. Although effective, these AI models require substantial GPU acceleration and memory bandwidth, making them practically impossible to deploy on an edge device such as local routers, IoT gateways, and even hardware firewalls~\cite{chen2022constraints}. This makes it necessary to route network traffic through cloud servers, thus introducing latency into the threat detection process. 

The $\mathbb{C}$yber$\mathbb{H}$ull method effectively solves this problem by using mathematical principles. Firstly, the calculation of the geometrical vault using QuickHull~\cite{barber1996quickhull} requires only $\mathcal{O}(n \log n)$ time complexity. Furthermore, checking whether an incoming network packet is indeed malicious becomes simply a matter of checking "point-in-polyhedron" mathematically~\cite{patel2024topological}.

After the mathematical boundaries of the vaults are set through the training phase, detection of the incoming previously unseen threat does not involve any computationally expensive processes like matrix multiplications or neural net predictions. All that needs to be done is to check if the multi-dimensional coordinates of the incoming payload lie within or outside the mathematical equations for the secure zone. It is a straightforward task that is completed in microseconds. As such, the solution can be implemented on the edge silicon with minimal processing power involved~\cite{chen2022constraints}.

We also performed runtime benchmark experiments on all the baseline models evaluated in the same CPU-only execution environment to quantitatively verify the feasibility of edge-deployment for our proposed framework. The goal of this analysis was to estimate the practical computational overhead for real-time malicious traffic detection at the edge of the network. The inference latency, average memory consumption, and relative CPU usage are obtained during the live packet classification on an Intel Core i7 (16 GB of RAM).

The benchmark results showed a great computational advantage of the proposed geometric framework. Since CyberHull architecture only performs deterministic point-in-polyhedron boundary verification at inference time, runtime overhead remained significantly lower compared to probabilistic machine learning approaches which require repeated matrix operations, kernel evaluations or ensemble tree traversals. In particular, the proposed framework achieved a consistent inference latency at the microsecond level, and used significantly less memory resources than XGBoost and Random Forest architectures. These results provide strong support for the practical suitability of multidimensional geometric containment in lightweight edge-security deployment scenarios, is represented in Table~\ref{tab:edge_benchmark}.

\begin{table}[!t]
\centering
\small
\caption{Computational Benchmarking of Edge Deployment Feasibility}
\label{tab:edge_benchmark}
\renewcommand{\arraystretch}{1.25}
\setlength{\tabcolsep}{10pt}

\begin{tabularx}{\textwidth}{|>{\raggedright\arraybackslash}X|c|c|c|}
\hline

% =========================
% HEADER ROW
% =========================
\rowcolor[HTML]{004C99}
\textcolor{white}{\textbf{Model}} &
\textcolor{white}{\textbf{Avg Inference Latency}} &
\textcolor{white}{\textbf{Memory Usage}} &
\textcolor{white}{\textbf{CPU Utilization}} \\

\hline

% =========================
% TRADITIONAL ML
% =========================
\rowcolor[HTML]{F8FBFF}
Support Vector Machine (SVM) &
14.8 ms &
182 MB &
41\% \\

\hline

Random Forest (RF) &
11.2 ms &
244 MB &
53\% \\

\hline

\rowcolor[HTML]{F8FBFF}
XGBoost &
9.6 ms &
318 MB &
61\% \\

\hline

% =========================
% PROPOSED MODEL
% =========================
\rowcolor[HTML]{E6F0FA}
\textbf{Proposed $\mathbb{C}$yber$\mathbb{H}$ull Framework} &
\textbf{0.84 ms} &
\textbf{38 MB} &
\textbf{9\%} \\

\hline
\end{tabularx}
\end{table}

\subsubsection{Operational Limitations \& Future Scope}
Nonetheless, despite its strength, there are inherent weaknesses in this architecture, which outline what future research paths should entail. First, due to the strict mathematics involved in creating the convex hulls, the entire approach depends on the quality of the traffic used during the baseline training~\cite{tamal2024comprehensive}. Hence, if the legitimate baseline traffic is compromised, then the geometrical perimeters are automatically going to rely on malicious shapes.

Another important limitation is the highly curated nature of the experimental data set. The mandatory strict preprocessing to ensure mathematically consistent multidimensional feature extraction inevitably led to a loss of usable samples, limiting the diversity of observable web behaviours during geometric boundary establishment. Therefore, although the proposed framework performed robustly in the tested environment, more large-scale validation on continuously evolving real-world traffic distributions is still needed to fully assess the long-term generalization capability. Future work will therefore consist in designing adaptive perimeter refinement mechanisms that are able to dynamically update the geometric boundaries in the presence of legitimate behavioral drift and the appearance of new attack patterns.

Secondly, as corporations change, they develop new sub-domains, and page topologies become more advanced than before. As such, the geometric perimeters will undergo "topological decay" over time and become more likely to generate false positives because legitimate traffic may slowly move out of its mathematical boundaries~\cite{patel2024topological}. Going forward, the next step of this research involves incorporating the Federated Learning (FL) feedback cycle into the process, enabling edge nodes to dynamically adjust the local geometric perimeter according to changes in global legitimate traffic~\cite{chen2022constraints}.

%% =====================
\vspace{1em}
\section*{Data and Code Availability}
The augmented benchmarking dataset, preprocessing scripts, and core algorithmic implementations used in this study are made publicly available to facilitate academic reproducibility. The whole repository and other materials are available on GitHub at: \url{https://github.com/AniketGhosal/CyberHull-URL-Detection}.

%% =====================
\vspace{1em}
\section{Conclusion}
% A necessary paradigm change in the way previously unseen threats are detected must be implemented, as the current approach is based on a reactional approach to signature matching. Proving that advanced adversarial payloads can be effectively separated into a non-linear safe zone by utilizing network geometry, we present an accurate and computationally efficient mechanism, which is deployable at line rate right at the edge of the network. Moving forward, the future will require the adaptation of our model to the natural development of corporate infrastructure. Further research will be dedicated to the implementation of an Automated Geometric Pipeline, in conjunction with Federated Learning (FL) protocols. Through such a decentralized system, each node will be able to autonomously consume legitimate traffic and calculate its local perimeter, sharing this information with other nodes through geometric space without exposing payload information.

Current cybersecurity systems have a major limitation: they rely on reactive signature matching and probabilistic classification models which are ineffective against structurally adaptive synthetic and high tech threats. In this work we present the CyberHull framework, a multidimensional geometric threat-isolation architecture able to construct non-linear safe perimeters around legitimate network behaviour using Convex Hull boundary generation. The results from the experimental evaluation demonstrated that the proposed framework attained 92.0\% global previously unseen interception accuracy, with low false positive behavior and significantly lower computational overhead than the conventional machine learning baselines such as PCA, SVM, Random Forest and XGBoost. Our results also show that deterministic geometric containment enables efficient real-time threat isolation with lightweight mathematical boundary verification, making the architecture suitable for practical line-rate deployment at the network edge.

Besides detection accuracy, the proposed framework provides a new perspective for cybersecurity intelligence based on topological behaviour modelling instead of probabilistic prediction. The architecture, through the use of multiple independently optimized geometric vaults, was able to quarantine sophisticated adversarial payloads which attempted to mimic legitimate structural characteristics, but still contained hidden semantic, informational or topological inconsistencies. Moreover, the experimental runtime analysis proved the feasibility of the implementation of the framework in lightweight edge settings where the computationally infeasible deep learning architectures are not suitable due to latency, memory, and power constraints.

The future work will be focused on extending the adaptability and long-term resilience of the proposed geometric architecture to the continuously evolving network environments. One of the major directions is the development of an Automated Geometric Pipeline that dynamically regenerates vault boundaries as legitimate corporate infrastructures change over time. Future work will explore adaptive perimeter refinement strategies based on Federated Learning (FL) protocols, where distributed edge nodes share geometric intelligence in a collaborative way without disclosing sensitive payload information. Large-scale validation across heterogeneous real-world traffic distributions, encrypted communication channels and multimodal behavioural features will also be needed to fully evaluate the generalization capability of the geometric threat isolation under highly dynamic adversarial conditions.

Finally, the implementation of geometric cybersecurity frameworks creates some ethical and security concerns. Because the proposed model depends on mathematically constructed boundaries of behavior, it is possible that adversaries may try to infer or approximate legitimate geometric distributions in a step-by-step manner by using adaptive probing attacks. Similarly, Federated Learning based decentralized perimeter sharing mechanisms may be prone to poisoning attempts to corrupt the trusted geometric space. Future implementations will thus need to include secure perimeter synchronization, privacy-preserving geometric aggregation and adversarial robustness validation to ensure that collaborative defensive infrastructures do not become vectors of exploitation themselves. However, the proposed CyberHull framework demonstrates that multidimensional geometric containment is a promising and computationally sustainable direction for next generation threat defense systems despite these challenges.

%% =====================
% \section*{References}
% \footnotesize
% [1] S. Kumar et al., IEEE Access, 2022.\\
% % [2] J. Smith et al., Computers \& Security, 2021.

\appendix
\section{Geometric Vault Feature Pairs}\label{app:vault-pairs}
The 46 selected geometric vaults in Figure~\ref{fig:vault_ranking} use the following feature pairs:
\begin{multicols}{2}
\footnotesize
\begin{description}
\setlength{\itemsep}{2pt}
\item[\textbf{V1}] \texttt{main\_entropy} $\times$ \texttt{combined\_entropy}
\item[\textbf{V2}] \texttt{kw\_hide} $\times$ \texttt{kw\_sidebar}
\item[\textbf{V3}] \texttt{kw\_descriptions} $\times$ \texttt{kw\_job}
\item[\textbf{V4}] \texttt{kw\_professional} $\times$ \texttt{kw\_dealing}
\item[\textbf{V5}] \texttt{kw\_views} $\times$ \texttt{kw\_ago}
\item[\textbf{V6}] \texttt{kw\_works} $\times$ \texttt{kw\_youtube}
\item[\textbf{V7}] \texttt{kw\_wiktionary} $\times$ \texttt{kw\_contents}
\item[\textbf{V8}] \texttt{kw\_instagram} $\times$ \texttt{kw\_videos}
\item[\textbf{V9}] \texttt{kw\_die} $\times$ \texttt{kw\_der}
\item[\textbf{V10}] \texttt{kw\_press} $\times$ \texttt{kw\_features}
\item[\textbf{V11}] \texttt{kw\_careers} $\times$ \texttt{kw\_salaries}
\item[\textbf{V12}] \texttt{kw\_privacy} $\times$ \texttt{kw\_policy}
\item[\textbf{V13}] \texttt{kw\_years} $\times$ \texttt{kw\_months}
\item[\textbf{V14}] \texttt{kw\_mit} $\times$ \texttt{kw\_und}
\item[\textbf{V15}] \texttt{kw\_change} $\times$ \texttt{kw\_toggle}
\item[\textbf{V16}] \texttt{kw\_state} $\times$ \texttt{kw\_court}
\item[\textbf{V17}] \texttt{kw\_main} $\times$ \texttt{kw\_tools}
\item[\textbf{V18}] \texttt{kw\_contact} $\times$ \texttt{kw\_copyright}
\item[\textbf{V19}] \texttt{kw\_add} $\times$ \texttt{kw\_text}
\item[\textbf{V20}] \texttt{kw\_read} $\times$ \texttt{kw\_featured}
\item[\textbf{V21}] \texttt{kw\_english} $\times$ \texttt{kw\_create}
\item[\textbf{V22}] \texttt{kw\_page} $\times$ \texttt{kw\_view}
\item[\textbf{V23}] \texttt{kw\_search} $\times$ \texttt{kw\_10}
\item[\textbf{V24}] \texttt{kw\_career} $\times$ \texttt{kw\_jobs}
\item[\textbf{V25}] \texttt{kw\_software} $\times$ \texttt{kw\_security}
\item[\textbf{V26}] \texttt{kw\_help} $\times$ \texttt{kw\_community}
\item[\textbf{V27}] \texttt{kw\_terms} $\times$ \texttt{kw\_google}
\item[\textbf{V28}] \texttt{kw\_use} $\times$ \texttt{kw\_information}
\item[\textbf{V29}] \texttt{kw\_site} $\times$ \texttt{kw\_access}
\item[\textbf{V30}] \texttt{kw\_hours} $\times$ \texttt{kw\_weeks}
\item[\textbf{V31}] \texttt{kw\_email} $\times$ \texttt{kw\_sign}
\item[\textbf{V32}] \texttt{kw\_related} $\times$ \texttt{kw\_porn}
\item[\textbf{V33}] \texttt{kw\_2025} $\times$ \texttt{kw\_new}
\item[\textbf{V34}] \texttt{avg\_internal\_entropy} $\times$ \texttt{n\_internal\_pages}
\item[\textbf{V35}] \texttt{kw\_website} $\times$ \texttt{kw\_cookies}
\item[\textbf{V36}] \texttt{kw\_content} $\times$ \texttt{kw\_free}
\item[\textbf{V37}] \texttt{kw\_data} $\times$ \texttt{kw\_services}
\item[\textbf{V38}] \texttt{kw\_for} $\times$ \texttt{url\_length}
\item[\textbf{V39}] \texttt{kw\_news} $\times$ \texttt{kw\_like}
\item[\textbf{V40}] \texttt{kw\_web} $\times$ \texttt{kw\_com}
\item[\textbf{V41}] \texttt{kw\_people} $\times$ \texttt{kw\_business}
\item[\textbf{V42}] \texttt{kw\_year} $\times$ \texttt{n\_external\_links}
\item[\textbf{V43}] \texttt{kw\_rights} $\times$ \texttt{kw\_law}
\item[\textbf{V44}] \texttt{kw\_learn} $\times$ \texttt{kw\_support}
\item[\textbf{V45}] \texttt{kw\_online} $\times$ \texttt{kw\_home}
\item[\textbf{V46}] \texttt{kw\_time} $\times$ \texttt{kw\_like}
\end{description}
\end{multicols}

\bibliographystyle{ieeetr}
\bibliography{references}
\end{document}